\documentclass[twocolumn,superscriptaddress,amsmath,amssymb,aps,prl,floatfix]{revtex4-2}
\usepackage[usenames,dvipsnames]{color}
\usepackage{xcolor}
\usepackage{multibib} 

\usepackage{comment}
\usepackage{cancel}
\usepackage{ulem}
\usepackage{physics}
\usepackage{graphicx}
\usepackage{dcolumn}
\usepackage{multirow}
\usepackage{bm}
\usepackage{gensymb}
\usepackage{tensor}
\usepackage[percent]{overpic}
\usepackage{tikz}
\usepackage{pgfplots}
\pgfplotsset{compat=1.15}
\usepackage{mathrsfs}
\usetikzlibrary{arrows}
\definecolor{qqttcc}{rgb}{0,0.2,0.8}
\definecolor{ccqqqq}{rgb}{0.8,0,0}
\definecolor{uuuuuu}{rgb}{0.26666666666666666,0.26666666666666666,0.26666666666666666}

\usepackage[breaklinks=true,colorlinks,citecolor=blue,linkcolor=blue,urlcolor=blue]{hyperref}
\definecolor{1}{HTML}{386CB0}
\definecolor{2}{HTML}{2CA02C} 

\begin{document}
\title{Optical properties, plasmons, and orbital Skyrme textures in twisted TMDs}
\date{\today}
\author{Lorenzo Cavicchi$^*$$^\dagger$}
\affiliation{Scuola Normale Superiore, I-56126 Pisa,~Italy}
\affiliation{Dipartimento di Fisica dell'Universit\`a di Pisa, Largo Bruno Pontecorvo 3, I-56127 Pisa, Italy}
\author{Koen J. A. Reijnders$^*$}
\affiliation{Radboud University, Institute for Molecules and Materials,
Heyendaalseweg 135, 6525 AJ Nijmegen,~The Netherlands}
\author{Mikhail I. Katsnelson}
\affiliation{Radboud University, Institute for Molecules and Materials, 
Heyendaalseweg 135, 6525 AJ Nijmegen,~The Netherlands}
\author{Marco Polini}
\affiliation{Dipartimento di Fisica dell'Universit\`a di Pisa, Largo Bruno Pontecorvo 3, I-56127 Pisa, Italy}
\affiliation{ICFO-Institut de Ci\`{e}ncies Fot\`{o}niques, The Barcelona Institute of Science and Technology, Av. Carl Friedrich Gauss 3, 08860 Castelldefels (Barcelona),~Spain}
\begin{abstract}
In the long-wavelength limit, Bloch-band Berry curvature has no effect on the bulk plasmons of a two-dimensional electron system. In this Letter we show instead that bulk plasmons are a probe of real-space topology. In particular, we focus on orbital Skyrme textures in twisted transition metal dichalcogenides, presenting detailed semiclassical and quantum mechanical calculations of the optical conductivity and plasmon spectrum of twisted ${\rm MoTe}_2$.
\end{abstract}

\maketitle

{\color{blue} {\it Introduction}.}---Forty years ago~\cite{berry_1984} Michael Berry introduced what are now routinely called ``Berry phase'' and ``Berry curvature''~\cite{Xiao_RMP_2010,Bernevig_Hughes,vanderbilt}. Soon after~\cite{Zak_PRL_1989}, Joshua Zak realized that these geometric concepts could be generalized to Bloch-periodic systems, where the parameters (quasi-momenta) are varied in closed loops (bands or Fermi surfaces) by applying electric fields.

Berry phase and Berry curvature have been powerful tools for understanding a plethora of intrinsic (i.e.~geometric) contributions to crystals' properties. These properties of the {\it bulk} Bloch bands have been identified to play a pivotal role in a wide range of physical phenomena, including electric polarization~\cite{resta_1992}, magnetic oscillations in metals~\cite{Mikitik_PRL_1999,Novoselov_Nature_2005,Zhang_Nature_2005}, anomalous Hall effects~\cite{Jungwirth_PRL_2002,Yao_PRL_2004,Nagaosa_RMP_2010}, orbital magnetism~\cite{xiao_2005}, quantum Hall effects of various kinds~\cite{laughlin_1981,thouless_1982,haldane_1988,kane_2005a,kane_2005b,fu_2006}, and quantized charge pumping~\cite{niu_1990}. Additionally, they are crucial also in systems where nontrivial {\it real space} (as opposed to momentum space) topology emerges,~e.g. when spin skyrmion lattices are present~\cite{bruno_PRL_2004,nagaosa_2013}. Originally proposed in nuclear physics by Skyrme in 1962~\cite{skyrme_1962}, skyrmions have become fundamental in understanding spin structures in condensed matter systems. Initially conceptualized as vortices in the spatial distribution of spin magnetization in crystals~\cite{bogdanov_1989}, skyrmions have been experimentally observed in magnetic materials~\cite{muhlbauer_science_2009, yu_nature_2010,yu_NatureMater_2010} and other physical systems, including focused photonic vector beams~\cite{du_2019} and suitably-designed space-coiling metastructures for localized spoof plasmons~\cite{deng_2022}. 
\begin{figure}[h!]
\centering
    \begin{overpic}[width=1\columnwidth]{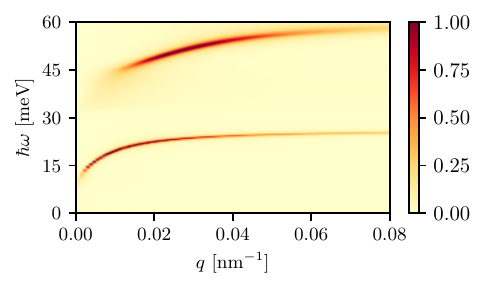}%
    \put(0,58){(a)}
    \end{overpic}\\
    \vspace{3mm}
    \begin{overpic}[width=1\columnwidth]{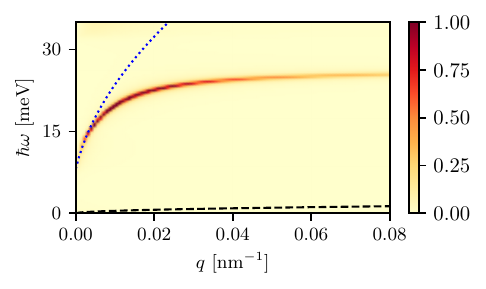}
    \put(0,58){(b)}
    \end{overpic}\\
    \vspace{0mm}
\caption{(Color online) The energy loss function ${\cal L}({\bm q},\omega)$ of twisted MoTe$_2$ as a function of the in-plane wave vector ${\bm q}$ and frequency $\omega$. Results in this plot refer to filling factor $\nu=-1$ and temperature $T= 5~{\rm K}$ (chemical potential $\mu \approx 41~{\rm meV}$). The twist angle is fixed at $\theta = 3.1^\circ$. Panel (a) shows the loss function for energies up to $60$ meV. Note that the lowest-energy mode, which is a slow inter-band plasmon, is gapped. The long-wavelength gap measures the ${\bm q}={\bm 0}$ uniform component of the Skyrme pseudo-magnetic field $B^z_{\rm eff}({\bm r})$ defined in Eq.~(\ref{eqn:magnetic-field-z}). Panel (b) shows the inter-band mode between the first and second valence bands. The dashed black line represents a thermally activated intra-band plasmon (Eq.~\eqref{eqn:intra-band_plasmon}). The dotted blue line is an analytical approximation for the gapped inter-band plasmon---see Eq.~\eqref{eqn:plasmon_dispersion}.\label{fig:fig1}}
\end{figure}

In this Letter, we focus on a different solid-state platform where skyrmion lattices play a crucial role, i.e.~twisted transition metal dichalcogenide (TMD) bilayers, such as twisted ${\rm MoTe}_2$ homobilayers. These materials have recently attracted a great deal of attention because of the experimental discovery of fractional Chern insulating states in zero magnetic field~\cite{Cai2023,Park2023,Zeng2023,Xu2023} and, more recently, superconductivity~\cite{xia_arxiv_2024,guo_arxiv_2024}. Available continuum-model Hamiltonians~\cite{wu_PRL_2019,pan_2020} describing their single-particle topological moir\'e bands---which have been recently dubbed {\it skyrmion Chern-band models}~\cite{morales-duran_PRL_2024,reddy_arxiv_2024}---harbor orbital (rather than spin) skyrmion lattices.

Intriguingly, because of the presence of this orbital skyrmion lattice, it is possible to approximately map these models into the problem of Landau levels subject to a periodic potential~\cite{morales-duran_PRL_2024,shi_PRB_2024}. More precisely, by means of an adiabatic approximation on the layer-pseudospin degree of freedom, the twisted TMD Hamiltonian transforms into one for holes under the effect of a periodic potential and a periodic magnetic field~\cite{morales-duran_PRL_2024,shi_PRB_2024}. This effective periodic magnetic field has a non-zero average, and its strength is related to the topological charge of the effective skyrmion lattice~\cite{wu_PRL_2019,pan_2020, morales-duran_PRL_2024,shi_PRB_2024}.
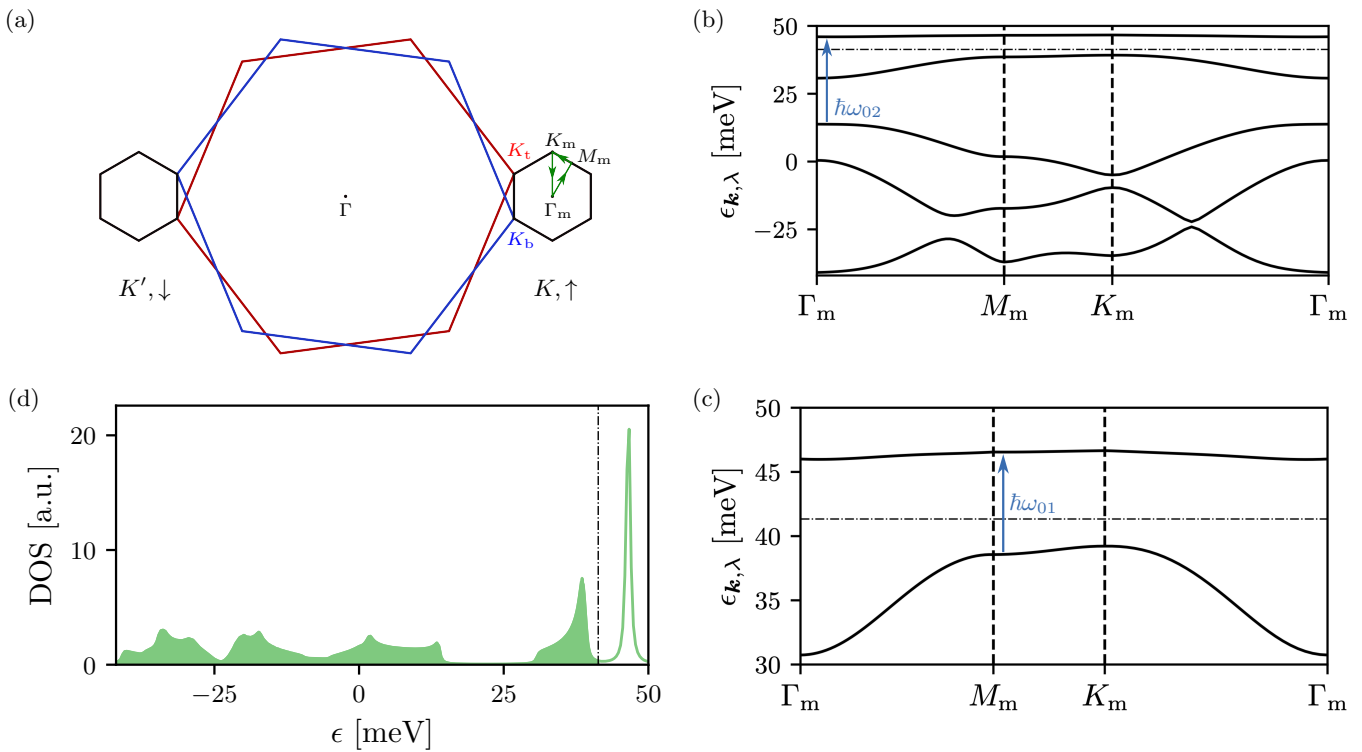
\begin{figure*}[!t]
\begin{tabular}{cc}
\begin{overpic}[width=0.4\textwidth]{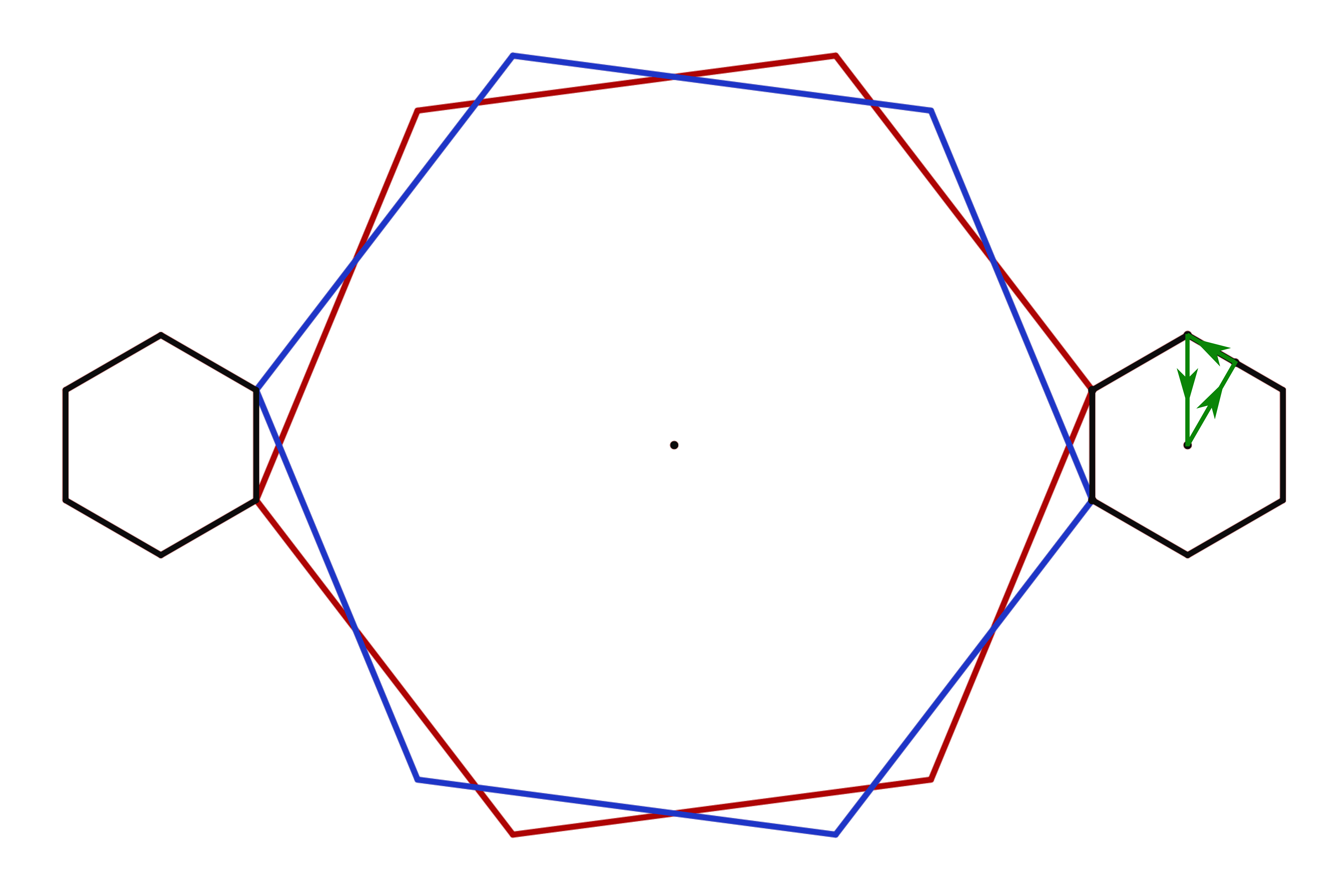}
\put(-13,65){(a)}%
\put(49,30){\scriptsize $\Gamma$}%
\put(87,30){\scriptsize $\Gamma_{\rm m}$}%
\put(87,43){\scriptsize $K_{\rm m}$}%
\put(80,41){{\color{red} \scriptsize $K_{\rm t}$}}%
\put(80,24.5){{\color{blue}\scriptsize $K_{\rm b}$}}%
\put(93,40){\scriptsize $M_{\rm m}$}%
\put(85,15){$K,\uparrow$}%
\put(8,15){$K^\prime,\downarrow$}%
\end{overpic}
 & \begin{overpic}[width=0.5\textwidth]{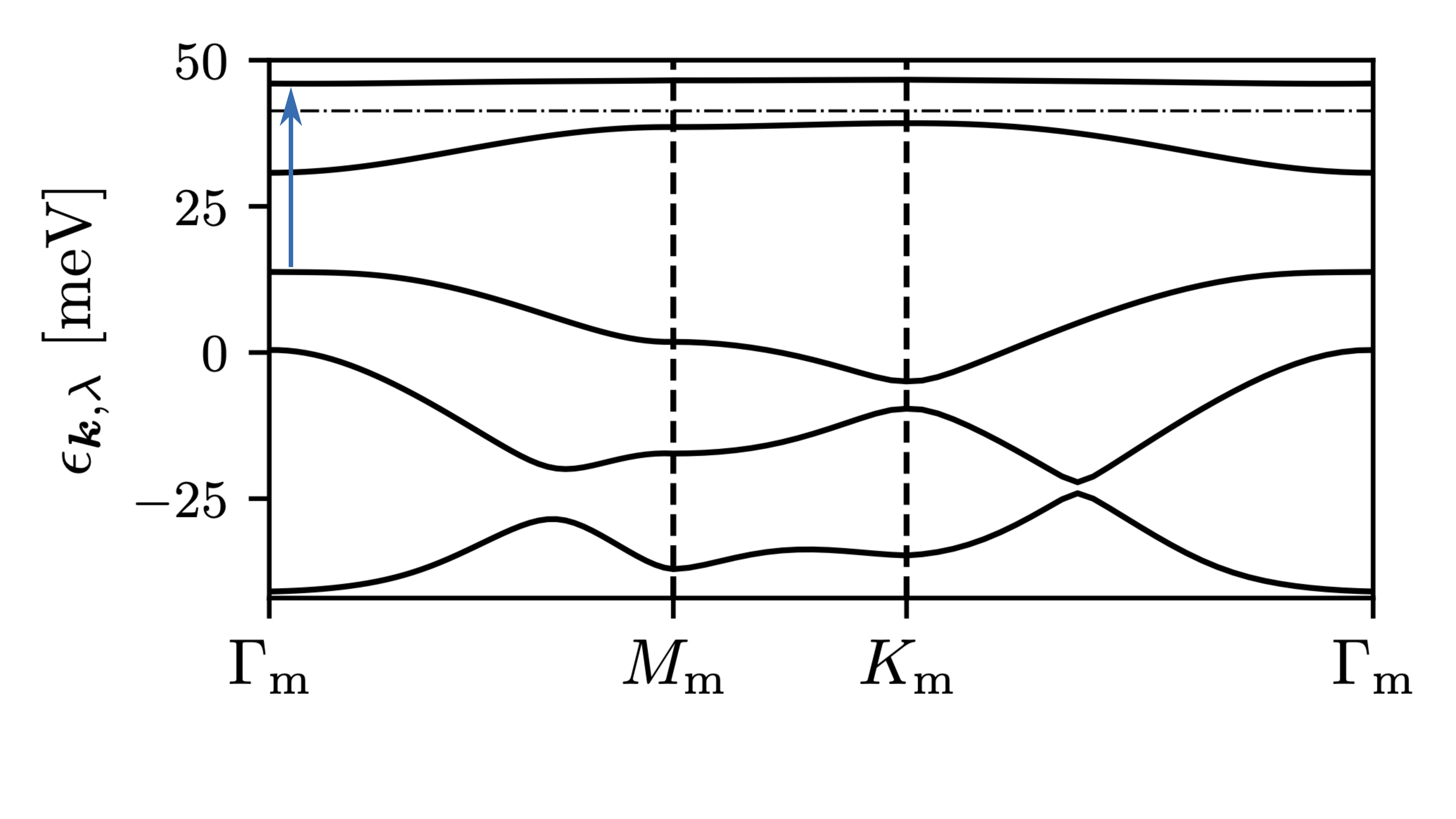}%
    \put(0,52){(b)}
    \put(21,39){\color{1}$\hbar\omega_{02}$}
    \end{overpic} \\ 
    \begin{overpic}[width=0.5\textwidth]{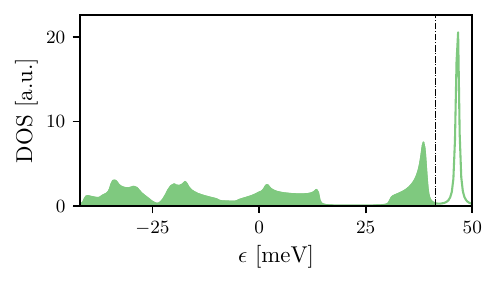}
\put(0,54){(d)}%
\end{overpic} & \begin{overpic}[width=0.5\textwidth]{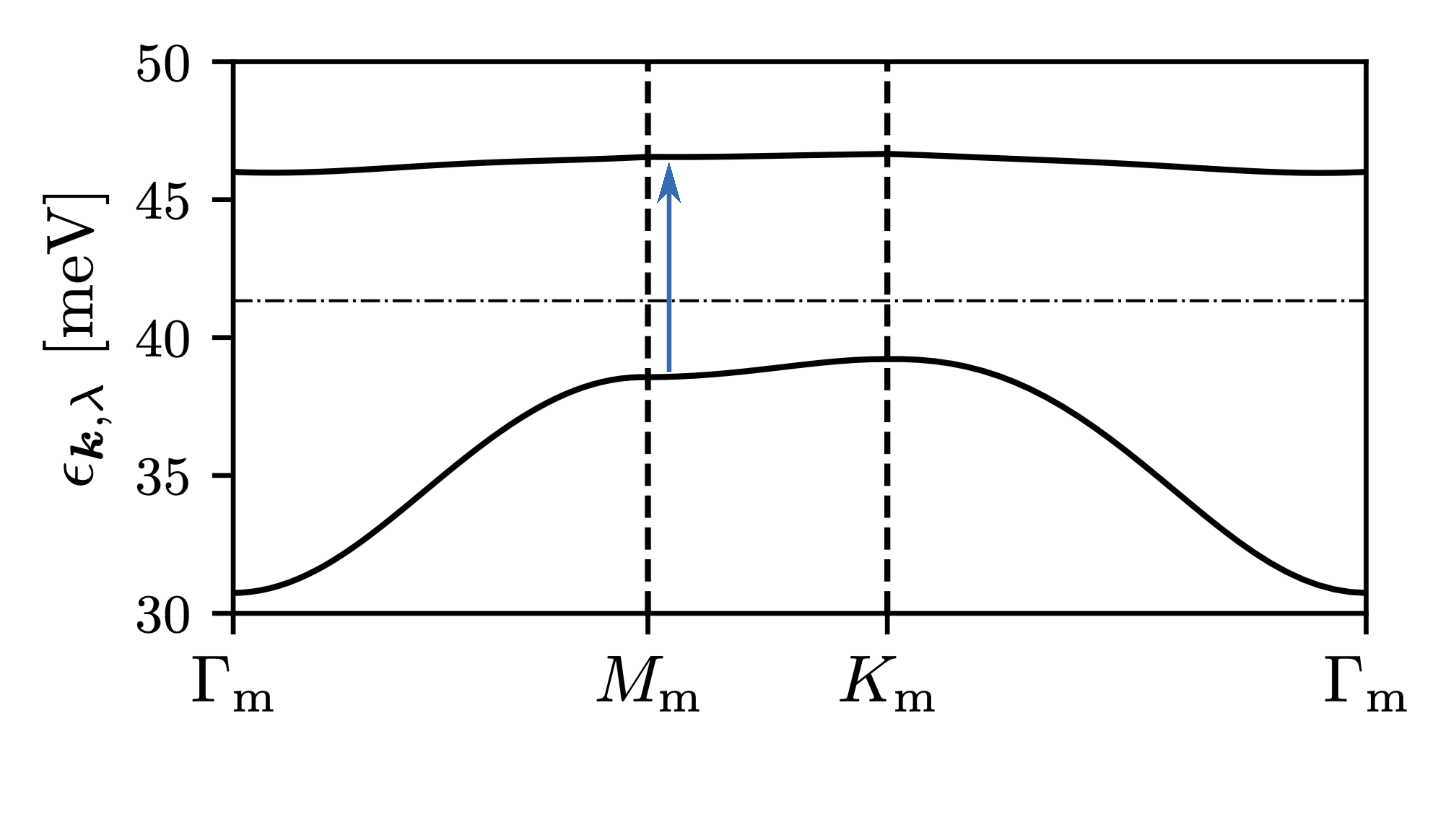}%
    \put(0,54){(c)}
    \put(47,38.5){\color{1}$\hbar\omega_{01}$}
    \vspace{5cm}
    \end{overpic}
\end{tabular}
\caption{Panel (a) The Brillouin zones (BZs) of two monolayer TMDs rotated relative to each other (blue and red big hexagons) are drawn together with the moir\'e BZ of a twisted homobilayer TMD (small black hexagons), constructed around the $K$ and $K^\prime$ valleys. The high symmetry path $\Gamma_{\rm m}$-$M_{\rm m}$-$K_{\rm m}$-$\Gamma_{\rm m}$ of the moir\'{e} BZ is highlighted by green arrows. Panel (b) The first five valence bands $\epsilon_{{\bm k}, \lambda}$ of twisted homobilayer ${\rm MoTe}_2$. The bands are obtained from   Eq.~\eqref{eqn:valence_hamiltonian}, for a twist angle $\theta = 3.1^\circ$. The dependence on ${\bm k}$ is displayed along the high symmetry path $\Gamma_{\rm m}$-$M_{\rm m}$-$K_{\rm m}$-$\Gamma_{\rm m}$ of the moir\'{e} BZ. Panel (c) shows a zoom of the first two valence bands. In panels (b) and (c),   blue vertical arrows connect pairs of nested bands, i.e.~mark interband transitions with the highest spectral weight. In panel (b), $\hbar\omega_{02}$ is between the first and third valence bands while, in panel (c), $\hbar\omega_{01}$  involves the first and second valence bands.  The energies associated with these transitions are: $\hbar\omega_{01}\approx 8~{\rm meV}$ and $\hbar\omega_{02}\approx 33~{\rm meV}$, respectively.  Panel (d) shows the density of states relative to the set of bands displayed in panel (b). The dashed dot line represents the chemical potential, fixed at $\mu \approx 41~{\rm meV}$, corresponding to a filling factor of $\nu = -1$ at $T=5$ K.\label{fig:fig2}}
\end{figure*}

In this Letter we first present a thorough theoretical study of the optical and plasmonic properties of these skyrmion Chern-band models. Despite numerical results are presented for the case of twisted ${\rm MoTe}_2$, the theoretical presentation is carried out in general. We then argue that plasmons in twisted TMD bilayers are a  probe of orbital skyrmion textures (see Fig.~\ref{fig:fig1}) in the sense that the gap at zero wave number is approximately related to the uniform component of the skyrmion effective magnetic field.
Our findings suggest that the effective magnetic field description can be used in order to qualitatively understand the main properties of this class of systems, while an accurate quantum treatment is instead needed for obtaining quantitative results. In addition, we perform a derivation of the mapping onto a single layer-pseudospin sector, which is alternative to the one of Refs.~\cite{morales-duran_PRL_2024,shi_PRB_2024} and based on semiclassical techniques. This gives rise to a formal series expansion, providing us with a systematic way of constructing all higher-order terms in the adiabatic parameter.

{\color{blue} {\it Bloch momentum-space topology has no effect on long-wavelength bulk plasmons}.}---Plasmons are collective excitations of the electron density in a crystal~\cite{GiulianiVignale}, and their behavior can be indeed influenced by various factors, including the underlying band structure and, in principle, its topological properties. To investigate the impact of Berry curvature on plasmons~\cite{song_PNAS_2016}, it is convenient to set up collisionless hydrodynamic equations, which describe the long-wavelength collective motion of the electron fluid. These can be then coupled to Hamilton equations of motion describing an electron wave packet in a crystal~\cite{xiao_2005}:
\begin{equation}\label{eqn:dot_r}
    \dot{\bm r} = \frac{1}{\hbar} \bm{\nabla}_{\bm k}\epsilon_{\lambda}(\bm k) + \bm{\Omega}_{\lambda}(\bm{k})\times\dot{\bm k}~,
\end{equation}
and
\begin{equation}\label{eqn:dot_k}
    \dot{\bm k} = -\frac{1}{\hbar} \bm{\nabla}_{\bm r} V(\bm r, t)~.
\end{equation}
Here, $\lambda$ is a band index, $\epsilon_{\lambda}(\bm k)$ is the band energy, $V(\bm r, t)$ is the instantaneous Hartree potential~\cite{GiulianiVignale}, and $\bm r, \bm k$ are the position and quasi-momentum of the Bloch wave packet, respectively. The second term on the right-hand-side of Eq.~\eqref{eqn:dot_r} is the so-called {\it anomalous velocity} and affects the motion of the wave packet when the band possesses a non-zero Berry curvature $\bm{\Omega}_{\lambda}(\bm{k})$, defined as:
\begin{equation}
\bm{\Omega}_{\lambda}(\bm{k}) \equiv i\langle\partial_{\bm k} u_{{\bm k},\lambda}|\times|\partial_{\bm k}u_{{\bm k},\lambda}\rangle~.
\end{equation}
Here, $|u_{{\bm k},\lambda}\rangle$ is the periodic part of the Bloch wave function corresponding to a given $\bm k$ and $\lambda$.

 We now write down the (collisionless) hydrodynamic equations describing plasmons, which are the Euler equation~\cite{GiulianiVignale},
\begin{equation}\label{eqn:euler}
    \partial_t \bm{j} + c_{\rm s}\bm{\nabla}_{\bm r}\delta n - \frac{e n_0}{m^*}\bm{\nabla}_{\bm r}V({\bm r}, t) = 0~,
\end{equation}
and the continuity equation,
\begin{equation}\label{eqn:continuity}
    \partial_t \delta n + \bm{\nabla}_{\bm r}\cdot \bm{j}_{\rm tot} = 0~.
\end{equation}
In Eq.~\eqref{eqn:euler}, $\bm{j} = \bm{p}/m^*$ is the current density, $c_{\rm s}$ is the sound velocity~\cite{GiulianiVignale}, $\delta n({\bm r},t) \equiv n({\bm r},t) - n_0$ is the deviation of the carrier density $n({\bm r},t)$ from the uniform density $n_0$ at equilibrium, and $m^*$ is the carrier effective mass. Notice that the total, {\it physical}  current density
\begin{equation}\label{eqn:physical_current}
    \bm{j}_{\rm tot} \equiv \bm{j} + \bm{\Omega}_{\lambda}(\bm{k})\times\dot{\bm k}
\end{equation}
appears in the continuity equation~\eqref{eqn:continuity}. Finally, $V({\bm r}, t)$ in Eq.~\eqref{eqn:euler} is related to the density fluctuation $\delta n$ through the Poisson equation:
\begin{equation}
    \nabla_{\bm r}^2 V({\bm r}, t) = 4\pi e \delta n({\bm r}, t)~,
\end{equation}
where, for the sake of simplicity, the dielectric environment has been set to the vacuum. 

Combining Eqs.~\eqref{eqn:physical_current} and~\eqref{eqn:dot_k}, we immediately conclude that the anomalous velocity contribution drops out of the continuity equation, since what matters there is the divergence of the physical current:
\begin{align}
\bm{\nabla}_{\bm r} \cdot \left[\bm{\Omega}_{\lambda}(\bm{k})\times\dot{\bm k}\right] &= -\frac{1}{\hbar}\bm{\nabla}_{\bm r} \cdot \left[\bm{\Omega}_{\lambda}(\bm{k})\times \bm{\nabla}_{\bm r} V(\bm r)\right]\nonumber\\
&= -\frac{1}{\hbar}\bm{\Omega}_{\lambda}(\bm{k}) \cdot \left[\bm{\nabla}_{\bm r}\times \bm{\nabla}_{\bm r} V(\bm r)\right] \nonumber\\
&= 0~.
\end{align}
This is one of the main results of the pioneering paper by Song and Rudner~\cite{song_PNAS_2016}, i.e.~in the long-wavelength (hydrodynamic) limit, a Bloch momentum-space Berry curvature does not modify the bulk plasmon dispersion. It does instead produce edge plasmons---which were dubbed {\it chiral Berry plasmons} by the authors of Ref.~\cite{song_PNAS_2016}---at the boundary between a topologically non-trivial phase and a trivial one.

In what follows, we study the impact of {\it real-space} topology---as induced by an orbital skyrmion lattice---on the optical and plasmonic properties of the hosting material. As mentioned above, we carry out these investigations in the realm of the continuum models that have been introduced~\cite{reddy_arxiv_2024,wu_PRL_2019, morales-duran_PRL_2024,shi_PRB_2024} to describe the topological moir\'e bands of twisted TMD homobilayers. In particular, we focus on the longitudinal and Hall conductivity and the energy loss function of twisted ${\rm MoTe}_2$. 

{\color{blue} {\it Skyrmion Chern-band models}.}---We consider a twisted homobilayer TMD constructed by stacking two TMD monolayers at a relative distance $d$ and rotating one with respect to each other by a twist angle $\theta$. Focusing attention on the $K$ valley and using a $2 \times 2$ representation in the layer-pseudospin degrees of freedom, the single-particle low-energy valence-band model Hamiltonian of such system is given by the following expression~\cite{wu_PRL_2019}:
\begin{align}\label{eqn:valence_hamiltonian}
    &\hat{\cal H}_{\uparrow,v}^{K}(\bm{k}) = \nonumber\\
    &\left(\begin{array}{cc}
        -\frac{(\hat{\bm p} - \hbar\bm{K}_{\rm b})^2}{2m^*} + \Delta_{{\rm b}}(\bm{r}) & \tilde{\Delta}_{\rm T}({\bm r}) \\
        \tilde{\Delta}_{\rm T}({\bm r})^\dagger & -\frac{(\hat{\bm p} - \hbar\bm{K}_{\rm t})^2}{2m^*} + \Delta_{{\rm t}}({\bm{r}})
    \end{array}\right)~,
\end{align}
where 
\begin{equation}\label{delta_T}
    \tilde{\Delta}_{\rm T}({\bm{r}}) \equiv w \left[1 + e^{i\bm{G}_2(\theta)\cdot\bm{r}} + e^{i\bm{G}_3(\theta)\cdot\bm{r}}\right]
\end{equation}
is an inter-layer potential and
\begin{equation}\label{delta_ell}
    \Delta_{\ell}(\bm{r}) \equiv 2V\sum_{j=1,3,5}\cos\left[\bm{G}_j(\theta)\cdot\bm{r}\pm\psi\right]
\end{equation}
is an intra-layer potential, the $+$ sign applying to $\ell = \rm t$ (top layer) and the $-$ sign to the $\ell = \rm b$ (bottom layer) case. These intra-layer potentials, which are periodic over the moir\'e unit cell, come from different alignments of the metal and chalcogen atoms of the top and bottom TMD monolayers~\cite{wu_PRL_2019}. The reciprocal lattice vectors $\bm{G}_j(\theta)$ appearing on the right-hand side of Eq.~\eqref{delta_ell} belong to the first shell of the moir\'e reciprocal lattice and are defined by
\begin{equation}
\bm{G}_j(\theta) = \frac{8\pi}{\sqrt{3}a}\sin\left(\frac{\theta}{2}\right)\left(\cos\left(j\frac{\pi}{3}\right), \sin\left(j\frac{\pi}{3}\right)\right)~.
\end{equation}
Here, $a$ is the lattice parameter of the underlying monolayer TMD of interest. In Eq.~\eqref{eqn:valence_hamiltonian}, the momentum shifts $\bm{K}_{\ell}$ correspond to the position of the $K$ valley of the top and bottom layers, after rotation---see Fig.~\ref{fig:fig2}(a):
\begin{equation}\label{eqn:K_ell}
    \bm{K}_{\ell} = \frac{8\pi}{\sqrt{3}a}\sin\left(\frac{\theta}{2}\right) \left(-\frac{1}{2}, \pm \frac{\sqrt{3}}{2}\right)~.
\end{equation}
As in the case of Eq.~(\ref{delta_ell}), the $+$ sign applies to $\ell = \rm t$ while the $-$ sign to the $\ell = \rm b$ (bottom layer) case.

Since TMD monolayers display, as a consequence of strong spin-orbit coupling, spin-valley locking~\cite{xiao_PRL_2012}, the single-particle valence-band Hamiltonian relative to the other valley, $K^\prime$, is obtained from $\hat{\cal H}_{\uparrow,v}^{K}(\bm{k})$ in Eq.~\eqref{eqn:valence_hamiltonian} by applying to it the time-reversal operator~\cite{wu_PRL_2019}. In practice, this corresponds to changing the sign of the momentum shifts $\bm{K}_{\ell}$ in Eq.~\eqref{eqn:K_ell}, while simultaneously taking the complex conjugate of the Hamiltonian~\eqref{eqn:valence_hamiltonian}.

The energy bands $\epsilon_{\bm k,\lambda}$ can be easily found by diagonalizing the Hamiltonian, provided that, for a fixed value of the twist angle $\theta$, one chooses a physically sound set of parameters $\left(V, \psi, w\right)$. In Figs.~\ref{fig:fig2}(b) and (c) we have reported the first few valence bands of twisted homobilayer ${\rm MoTe}_2$, as obtained from Eq.~\eqref{eqn:valence_hamiltonian} at a twist angle $\theta = 3.1^{\circ}$, with the following choice of microscopic parameters~\cite{wang_prl_2024}: $a = 0.352~{\rm nm}$, $m^* = 0.6~m_{\rm e}$,  $m_{\rm e}$ being the bare electron mass in vacuum, and $\left(V, \psi, w\right) = (20.8~{\rm meV}, 107.7^\circ, -23.8~{\rm meV})$. The corresponding density of states has been reported in Fig.~\ref{fig:fig2}(d). 

We now comment on the origin of the name ``skyrmion Chern bands'' for the bands of the model defined by Eq.~(\ref{eqn:valence_hamiltonian}). The key point is that, upon carrying out a suitable gauge transformation~\cite{morales-duran_PRL_2024}, the valence-band Hamiltonian~\eqref{eqn:valence_hamiltonian} reduces to the following physically transparent form:
\begin{equation}\label{eqn:valence_hamiltonian2}
\hat{\cal H} = -\frac{\hat{\bm p}^2}{2m^*}\tau_0 + \bm{\Delta}(\bm{r})\cdot\bm{\tau} + \Delta_{0}(\bm{r})\tau_0~,
\end{equation}
where we have omitted the indices $\uparrow$ and $K$ and introduced the following quantities:
\begin{equation}\label{eqn:delta}
    \bm{\Delta}({\bm r}) \equiv \left(\Re \Delta_{\rm T}({\bm r}), \Im \Delta_{\rm T}({\bm r}), \frac{\Delta_{\rm b}({\bm r}) - \Delta_{\rm t}({\bm r})}{2}\right)~, 
\end{equation}
\begin{equation}\label{eqn:delta_0}
\Delta_0({\bm r}) \equiv \frac{\Delta_{\rm b}({\bm r}) + \Delta_{\rm t}({\bm r})}{2}~,
\end{equation}
and
\begin{equation}\label{eqn:delta_T}
    \Delta_{\rm T}(\bm{r}) \equiv w \left(e^{i\bm{q}_1\cdot\bm{r}} + e^{i\bm{q}_2\cdot\bm{r}} + e^{i\bm{q}_3\cdot\bm{r}}\right)~,
\end{equation}
with
\begin{equation}
    \bm{q}_1 = -\frac{8\pi}{\sqrt{3} a}\sin\left(\frac{\theta}{2}\right)\left(0,\frac{1}{\sqrt{3}}\right)~,
\end{equation}
$\bm{q}_2 = \bm{G}_2+\bm{q}_1$, and $\bm{q}_3 = \bm{G}_3+\bm{q}_1$. In Eq.~\eqref{eqn:valence_hamiltonian2}, the matrices ${\bm \tau}$ and $\tau_0$ are Pauli matrices acting on the layer-pseudospin (rather than spin) degrees of freedom.

The Hamiltonian in Eq.~\eqref{eqn:valence_hamiltonian2} is the analog of a spin-skyrmion Hamiltonian, with the skyrmion texture defined by $\bm{\Delta}(\bm{r})/|\bm{\Delta}(\bm{r})|$~\cite{wu_PRL_2019, morales-duran_PRL_2024}. The only difference is that real spin is here replaced by the {\it layer-pseudospin} degree of freedom.

\begin{figure*}[t]
\centering
\begin{tabular}{lll}
    \begin{overpic}[width=0.5\textwidth]{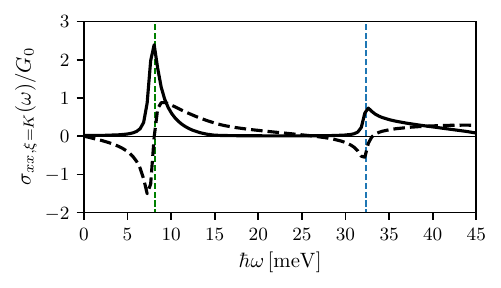}%
    \put(0,52){(a)}
    \put(32,16.5){\color{2}$\hbar\omega_{01}$}
    \put(74,16.5){\color{1}$\hbar\omega_{02}$}
    \end{overpic} & \begin{overpic}[width=0.5\textwidth]{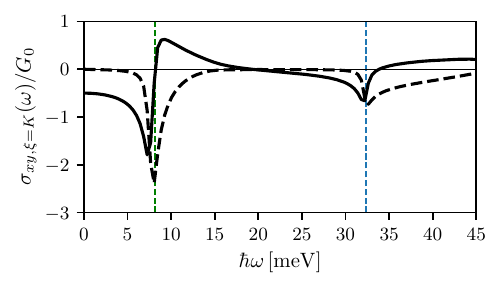}%
    \put(0,52){(b)}
    \put(32,16.5){\color{2}$\hbar\omega_{01}$}
    \put(74,16.5){\color{1}$\hbar\omega_{02}$}
    \end{overpic}
\end{tabular}
\caption{Single-valley and single-spin local optical conductivity $\sigma_{\alpha\beta, \xi=K}(\omega)$ of MoTe$_2$ as a function of $\omega$ in units of $G_0 = 2e^2/h$. Results in this plot refer to filling factor $\nu=-1$ and temperature $T= 5~{\rm K}$. The twist angle is fixed at $\theta = 3.1^\circ$. Panel (a) shows the longitudinal component $\sigma_{xx, K}(\omega)$; panel (b) shows the Hall component $\sigma_{xy, K}(\omega)$. In both panels the solid (dashed) line refers to the real (imaginary) part. The colored vertical dashed lines highlight the transition energies $\hbar\omega_{01}\approx 8~{\rm meV}$ (green) and $\hbar\omega_{02}\approx 33~{\rm meV}$ (blue).\label{fig:fig3}}
\end{figure*}
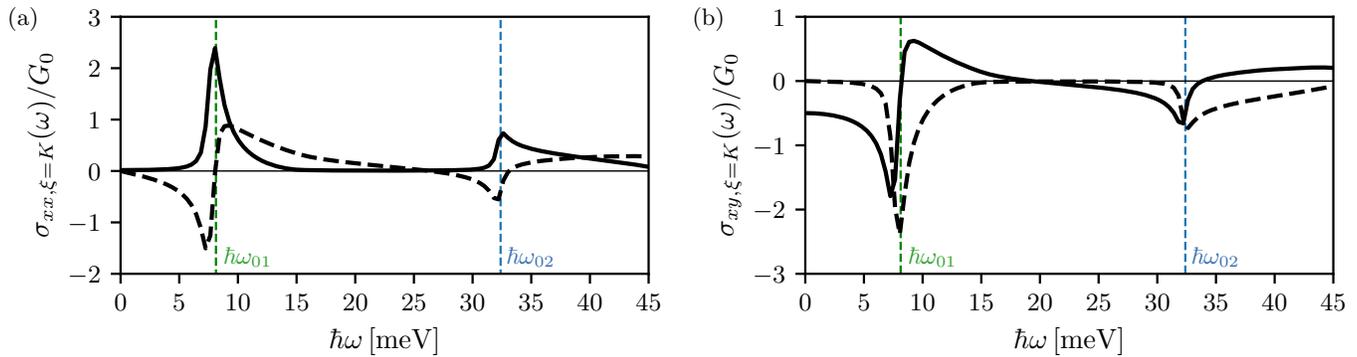

{\color{blue} {\it Local optical conductivity}.}---We now proceed to calculate the local optical conductivity tensor $\sigma_{\alpha\beta}(\omega)$ of a twisted homobilayer TMD. The quantity $\sigma_{\alpha\beta}(\omega)$ is the linear response function relating the electrical current flowing in the Cartesian direction $\alpha$ in response to a uniform electric field applied in the direction $\beta$~\cite{GiulianiVignale}. By using the Kubo formula~\cite{GiulianiVignale}, it can be shown that in a 2D crystal, i.e.~in a Bloch translationally-invariant system in which the single-particle eigenstates are of the Bloch type, $\sigma_{\alpha\beta}(\omega)$ is given by~\cite{novelli}
\begin{equation}\label{eqn:conductivity}
    \sigma_{\alpha\beta}(\omega) = \sum_{\xi = K,K^\prime} \sigma_{\alpha\beta,\xi}(\omega)~,
\end{equation}
where
\begin{equation}
    \sigma_{\alpha\beta,\xi}(\omega) = \sigma_{\alpha\beta,\xi}^{\rm intra}(\omega) + \sigma_{\alpha\beta,\xi}^{\rm inter}(\omega)
\end{equation}
is the sum of intra-band and inter-band contributions. Note that, in Eq.~(\ref{eqn:conductivity}), we have explicitly divided the {\it physical} optical conductivity $\sigma_{\alpha\beta}(\omega)$ into its valley-resolved contributions $\sigma_{\alpha\beta,\xi}(\omega)$, with $\xi = K,K^\prime$. We stress that, due the underlying giant spin-orbit coupling at the monolayer TMD level, also in the case of the twisted homobilayer TMD the spin degree of freedom is locked to the valley one. Therefore, the quantity $\sigma_{\alpha\beta,\xi}(\omega)$ physically represents a single-valley {\it and} single-spin, optical conductivity tensor.

The final expressions for the single-spin and single-valley intra- and inter-band optical conductivities are:
\begin{equation}\label{eqn:conductivity_intra}
    \sigma_{\alpha\beta,\xi}^{\rm intra}(\omega) = \frac{\hbar}{\pi}\frac{i{\cal D}_{\alpha\beta,\xi}}{\hbar\omega + i\eta}~,
\end{equation}
where
\begin{align}\label{eqn:drude_weight}
    {\cal D}_{\alpha\beta,\xi} =&{}
    \frac{\pi e^2}{\hbar^2} \sum_\lambda \int \frac{d^2\bm{k}}{(2\pi)^2}[-f_{\bm k, \lambda}^\prime]\times\nonumber\\
    &\times{}_\xi\langle \bm k, \lambda|\frac{\hbar}{m^*}\hat{p}_\alpha|\bm k, \lambda\rangle_\xi \langle \bm k, \lambda| \frac{\hbar}{m^*}\hat{p}_\beta|\bm k, \lambda\rangle_\xi ~,
\end{align}
is the Drude weight, and
\begin{align}\label{eqn:conductivity_inter-band}
\sigma_{\alpha\beta,\xi}^{\rm inter}(\omega) = &{}\frac{ie^2}{\hbar} \sum_{\lambda\neq\lambda^\prime} \int \frac{d^2\bm{k}}{(2\pi)^2}\left[-\frac{f_{\bm k, \lambda} - f_{\bm k, \lambda^\prime}}{\epsilon_{\bm k,\lambda}- \epsilon_{\bm k,\lambda^\prime}}\right]\times \nonumber\\
&\times \frac{{}_\xi\langle \bm k, \lambda|\frac{\hbar}{m^*}\hat{p}_\alpha|\bm k, \lambda^\prime\rangle_\xi \langle \bm k, \lambda^\prime| \frac{\hbar}{m^*}\hat{p}_\beta|\bm k, \lambda\rangle_\xi}{\epsilon_{\bm k,\lambda}- \epsilon_{\bm k,\lambda^\prime} + \hbar \omega + i \eta}~.
\end{align}
Here, $\epsilon_{\bm k,\lambda}$ is the eigenvalue associated with the Bloch eigenstate $|\bm k, \lambda\rangle_\xi$, $f_{\bm k, \lambda}$ is the Fermi-Dirac distribution, $f_{\bm k, \lambda}^\prime$ is its derivative with respect to the energy $\epsilon_{{\bm k}, \lambda}$, $\hat{p}_\alpha$ is the $\alpha$-th Cartesian component of the momentum operator, and $\eta =0^+$ is a positive infinitesimal. We point out that the integrals in Eqs.~\eqref{eqn:drude_weight} and~\eqref{eqn:conductivity_inter-band} are performed over the first moir\'e BZ. Energy bands $\epsilon_{\bm k,\lambda}$ and Bloch states $|\bm{k},\lambda\rangle_\xi$ are found numerically, by diagonalizing the Hamiltonian in Eq.~\eqref{eqn:valence_hamiltonian}.

Fig.~\ref{fig:fig3} shows the single-valley and single-spin optical conductivity of ${\rm MoTe}_2$ evaluated at a twist angle  $\theta  = 3.1^\circ$ and for an integer filling factor $\nu = -1$---see Fig.~\ref{fig:fig2}(d). The longitudinal part, which is displayed in Fig.~\ref{fig:fig3}(a), shows two peaks, the first one, at $\hbar\omega = \hbar\omega_{01} \approx 8~{\rm meV}$,  which is associated to an interband transition between the first two valence bands and the second one, at energy $\hbar\omega = \hbar\omega_{02}\approx33~{\rm meV}$, which is associated to an interband transition between the first and third valence bands. 

Fig.~\ref{fig:fig3}(b) shows instead the transverse {\it Hall-like} component of the optical conductivity tensor, which is nonzero in the single-valley and single-spin sector. In the following, we provide a physical interpretation of the appearance of such a finite Hall-like conductivity, showing that its origin stems from the orbital magnetic moment of the Bloch states~\cite{SM}. Another possibility to explain the finiteness of $\sigma_{xy, \xi}(\omega)$ at the level of the single-spin and single-valley model is to invoke a mapping into a Landau-level problem~\cite{morales-duran_PRL_2024}. In such a Landau-level mapping, $\sigma_{xy, \xi}(\omega)\neq 0$ is a direct consequence of the effective magnetic field due to the presence of the orbital skyrmion lattice. Finally, the fact that $\sigma_{xy,\xi} \neq 0$ ultimately stems from the broken $C_2{\cal T}$ symmetry (inversion times time-reversal)~\cite{Bernevig_Hughes}. This is due to the $z$ component of $\bm{\Delta}$~\cite{Bernevig_Hughes,bhowal_PRB_2021} in Eq.~\eqref{eqn:valence_hamiltonian2}, defined in Eq.~\eqref{eqn:delta}. This contribution allows for a nonzero orbital magnetic moment in each valley~\cite{bhowal_PRB_2021}. This is in stark contrast to the case of twisted bilayer graphene, which is  $C_2{\cal T}$ symmetric (and this is why Dirac points around the moir\'e $K$ and $K^\prime$ points are protected in this material) and $\sigma_{xy, \xi}\equiv 0$.

Despite the finiteness of $\sigma_{xy, \xi}$ in our study of twisted homobilayer TMDs, in the case in which time-reversal symmetry is not explicitly broken, the transverse Hall conductivity is odd, i.e.~$\sigma_{xy,K} = -\sigma_{xy,K^\prime}$. The total {\it physical} conductivity $\sigma_{\alpha\beta}$ defined in Eq.~\eqref{eqn:conductivity} is therefore purely longitudinal. Indeed, the longitudinal single-valley and single-spin optical conductivity is even under the exchange of the valley index, i.e.~$\sigma_{\alpha\alpha,K} = \sigma_{\alpha\alpha,K^\prime}$. 

Although the single-valley and single-spin optical conductivity calculations described in this Section are performed at the single-particle (i.e.~non-interacting) level, we expect that they qualitatively capture the physics of valley and spin-polarized states when electron-electron interactions are taken into account~\cite{wang_prl_2024}. Indeed, strong interactions tend to lift up the valley and spin degeneracies, leading to a flavor-polarized ground state~\cite{wang_prl_2024,reddy_PRB_2023,yu_PRB_2024}. Alternatively, since the skyrmion Chern band model~\eqref{eqn:valence_hamiltonian2} yields a finite Bloch orbital magnetic moment in each valley~\cite{SM}, a small magnetic field applied perpendicular to the electronic plane could break the valley degeneracy~\cite{xiao_PRL_2007}, allowing the probing of the single-valley and single-spin optical conductivity.

{\color{blue} {\it Collective modes: loss function}.}---We now turn to discuss collective modes in twisted homobilayer TMDs. In particular, we focus on the electron energy loss function ${\cal L}(\bm{q},\omega)$, which represents the probability of exciting the electron system by applying a scalar perturbation with wave vector $\bm{q}$ and energy $\hbar \omega$ that couples to the total electron density. This quantity therefore describes also self-sustained charge oscillations (plasmons), which appear as sharp peaks, and contains information about the spectral  density of incoherent electron-hole pairs, which yield a broadly distributed background in the ${\bm q}$-$\omega$ plane. The energy loss function can be measured in principle via electron energy loss spectroscopy~\cite{egerton} and scattering-type near-field optical spectroscopy~\cite{Basov_Nanophotonics_2021,Low_NatureMaterials_2017,Lundeberg_Science_2017,Woessner_ACS_2017,AlcarazIranzo_Science_2018,ni_nature_2018}. In crystals, it is defined as:
\begin{equation}\label{eqn:loss_function_xstal}
    {\cal L}(\bm{q},\omega) = -\Im\left\{\left[\varepsilon(\bm{q},\omega)^{-1}\right]_{\bm{G}=\bm{0},\bm{G}^\prime=\bm{0}}\right\}~,
\end{equation}
where $\varepsilon(\bm{q},\omega) = [\varepsilon(\bm{q},\omega)]_{{\bm G}, {\bm G}^\prime}$ is the dynamical dielectric function, which is a matrix~\cite{GiulianiVignale} with respect to the reciprocal lattice vectors $\bm{G}$ and $\bm{G}^\prime$. 

In the random phase approximation (RPA)~\cite{GiulianiVignale} and in the local limit (i.e.~$\bm{q}\to\bm{0}$ and $\bm{G},\bm{G}^\prime = \bm{0}$), this matrix can be expressed as~\cite{novelli}:
\begin{align}
\label{eqn:dielectric_tensor}
\lim_{\bm{q}\to\bm{0}}\left[\varepsilon(\bm{q}, \omega)\right]_{\bm{0},\bm{0}} \approx 1 + i\frac{L(\bm{q},\omega)}{\omega} q_\alpha q_\beta\sigma_{\alpha\beta}(\omega)~,
\end{align}
where $L(\bm{q},\omega)$ is the Coulomb propagator relating the charge density fluctuations $\delta n_{\bm{q}}(\omega)$ to the self-induced electrical potential, i.e.~$W(\bm{q},\omega) = e^2 L(\bm{q},\omega) \delta n_{\bm{q}}(\omega)$. Notice that, in Eq.~\eqref{eqn:dielectric_tensor}, the total physical optical conductivity \eqref{eqn:conductivity} appears. 

For a 2D system embedded in a homogeneous and isotropic dielectric environment with a dielectric permittivity $\bar{\varepsilon}$, the non-retarded Coulomb propagator is~\cite{keldysh}:
\begin{equation}  \label{eqn:Coulomb-propagator}
    L(\bm{q},\omega) \equiv \frac{2\pi}{q \bar{\varepsilon}}~.
\end{equation}
Note that we have neglected the frequency dependence of $\bar{\varepsilon}$ on purpose, with the aim of highlighting {\it intrinsic} features of the plasmonic spectrum of twisted homobilayer TMDs. Also, in our numerical calculations we have considered an isolated system in vacuum (i.e.~a suspended twisted homobilayer TMD), for which  $\bar{\varepsilon} = 1$. Extrinsic effects, which may modify the plasmonic spectrum, such as hyperbolic phonon polaritons~\cite{dai_SCIENCE_2014,caldwell_NAT_COMM_2014, li_NAT_COMM_2015} that appear e.g. when twisted homobilayer TMDs are encapsulated in hexagonal Boron Nitride (hBN), can be easily taken into account~\cite{tomadin_PRL_2015} but are not of our interest here.

A summary of our main results for the loss function ${\cal L}(\bm{q},\omega)$ of twisted homobilayer ${\rm MoTe}_2$ is reported in Fig.~\ref{fig:fig1}. Results in this figure have been obtained for the exact same choice of parameters as in Fig.~\ref{fig:fig3}. In the explored range of values of $\hbar \omega$, we clearly see two {\it sharp} and {\it slow} collective modes, the lower energy one falling in  the Terahertz spectral range. (Therefore this low energy mode will not be modified by the phonon polariton modes of the hBN slabs encapsulating ${\rm MoTe}_2$, which are at much higher energies, in the mid-infrared range.) The sharpness and slowness of these two collective modes is linked (as in the case of plasmons in twisted bilayer graphene~\cite{stauber_2016, lewandowski_PNAS_2019,novelli}) to the flatness of the moir\'e bands. Their existence can be inferred from the above mentioned single-particle optical inter-band transitions with highest spectral weight. Indeed, considering that we have chosen a hole density such that the filling factor is $\nu = -1$, the lowest-energy single-particle electron-hole excitation is the vertical inter-band transition between the first two flat valence bands, depicted as a blue arrow in Fig.~\ref{fig:fig2}(c), with associated transition energy $\hbar\omega = \hbar\omega_{01}\approx 8~{\rm meV}$. Upon including electron-electron interactions, as in our RPA calculations, these single-particle excitations merge into an inter-band collective plasmon mode, whose dispersion relation terminates at a finite value of $\hbar \omega$ given by $\hbar\omega = \hbar\omega_{01}$ in the long-wavelength $\bm{q}=\bm{0}$ limit. Similarly, the second collective mode occurring at higher energies, which is clearly visible in Fig.~\ref{fig:fig1}, stems from a coalescence of single-particle inter-band transitions involving the first and third valence bands, depicted as a blue arrow in Fig.~\ref{fig:fig2}(b), with an associated transition energy $\hbar\omega = \hbar\omega_{02}\approx 33~{\rm meV}$. Finally, the gapped behavior of these modes resembles that of {\it magnetoplasmons}~\cite{fetter_1985}, which are collective modes of 2D electron liquids subject to a strong perpendicular magnetic field. We will come back to this point below.

We now comment on the role of finite-temperature effects. Since our calculations are performed at finite $T$, we need, at least in principle, to check whether the results shown in Fig.~\ref{fig:fig1} can be explained in terms of a thermally activated intra-band plasmon. The latter would have a long-wavelength dispersion given by~\cite{novelli} 
\begin{equation}\label{eqn:intra-band_plasmon}
\omega_{\rm th}(q) = \sqrt{\frac{2{\cal D}_{\rm L}}{\bar{\varepsilon}} q}~,
\end{equation}
where ${\cal D}_{\rm L} \equiv {\cal D}_{xx, K} + {\cal D}_{xx, K^\prime} = {\cal D}_{yy, K} + {\cal D}_{yy, K^\prime}$ is the longitudinal part of the Drude weight tensor defined in Eq.~\eqref{eqn:drude_weight}. Such intra-band mode is depicted in Fig.~\ref{fig:fig1}(b) with a black dashed line. We conclude that our findings in Fig.~\ref{fig:fig1} cannot be explained in terms of trivial, thermally-activated, intra-band plasmons.

{\color{blue}{\it Interpretation of the numerical results}.}---In what follows, we provide a physical interpretation of our main numerical results in Fig.~\ref{fig:fig1} and~\ref{fig:fig3}(b). Starting from the Kubo formula for the optical conductivity, we first discuss the orbital character of the Hall-like component of the conductivity tensor and then provide a simple analytical formula for the gapped inter-band plasmon dispersion. 

Since the chemical potential lies in the gap between the first two flat valence bands and we work at relatively low temperatures ($k_{\rm B } T \ll \hbar\omega_{01}$), we neglect the intra-band contribution to the conductivity~\eqref{eqn:conductivity_intra} and concentrate only on the inter-band term in Eq.~\eqref{eqn:conductivity_inter-band}. In the analysis below, we further restrict our study to the contribution to $\sigma_{\alpha\beta,\xi}^{\rm inter}(\omega)$ stemming from the lowest inter-band transition at $\hbar \omega = \hbar\omega_{01}$.

Manipulating Eq.~\eqref{eqn:conductivity_inter-band} in the case of the Hall-like term $\sigma_{xy,\xi}^{\rm inter}(\omega)$ we find~\cite{SM}:
\begin{align}\label{eqn:orbital_conductivity}
    &\sigma_{xy,\xi}^{\rm inter}(\omega) =\nonumber\\
    &- ec \sum_{\lambda\neq\lambda^\prime} \int \frac{d^2\bm{k}}{(2\pi)^2}\frac{f_{\bm k, \lambda} - f_{\bm k, \lambda^\prime}}{\epsilon_{\bm k,\lambda}- \epsilon_{\bm k,\lambda^\prime}+ \hbar \omega + i \eta}\bm{m}_{\bm k,\lambda,\lambda^\prime}^\xi\cdot\bm{z}~,
\end{align}
where 
\begin{align}\label{eqn:bloch_orbital_momentum}
    &\bm{m}_{\bm k,\lambda,\lambda^\prime}^\xi \equiv \frac{ie}{2\hbar c}\times \nonumber \\
    &\frac{{}_\xi\langle u_{\bm{k},\lambda}|\partial_{\bm k} \hat{\cal H}^\xi(\bm{k}) |u_{\bm{k},\lambda^\prime}\rangle_\xi\times{}_\xi\langle u_{\bm{k},\lambda^\prime}|\partial_{\bm k} \hat{\cal H}^\xi(\bm{k}) |u_{\bm{k},\lambda}\rangle_\xi}{\epsilon_{\bm{k},\lambda} - \epsilon_{\bm{k},\lambda^\prime}}
\end{align}
is the inter-band orbital magnetic moment of the Bloch states~\cite{vanderbilt}. Here, $|u_{\bm{k},\lambda}\rangle_\xi$ denotes the periodic part of the Bloch state~\cite{SM}. Eq.~\eqref{eqn:orbital_conductivity} shows that a non-zero orbital magnetic moment of the Bloch states can result in a Hall-like conductivity $\sigma_{xy,\xi}^{\rm inter}(\omega)$. In our case, since we are dealing with a single-valley and single-spin model where the $C_2{\cal T}$ symmetry is broken, we obtain a non-trivial orbital magnetic moment texture in momentum space~\cite{vanderbilt, bhowal_PRB_2021}. In Ref.~\cite{SM} we further manipulate Eq.~\eqref{eqn:orbital_conductivity} in order to capture analytically the behavior of $\sigma_{xy,\xi}^{\rm inter}(\omega)$ near $\hbar \omega  = \hbar\omega_{01}$, and show numerical results for $\bm{m}_{\bm k,\lambda,\lambda^\prime}^\xi$ in the first moir\'{e} BZ. 

As far as the plasmonic modes are concerned, we stress that Eq.~\eqref{eqn:dielectric_tensor} is controlled only by the longitudinal part $\sigma_{xx}(\omega)$ of the physical optical conductivity since $\sigma_{xy}(\omega) = 0$ due a  cancellation of the two valley-resolved contributions $\sigma_{xy,\xi}(\omega)$. 
Imposing the condition $\lim_{\bm{q}\to\bm{0}}\left[\varepsilon(\bm{q}, \omega)\right]_{\bm{0},\bm{0}} = 0$ for the existence of a plasmon mode, we find the following analytical expression for the long-wavelength dispersion relation~\cite{SM}:
\begin{equation}\label{eqn:plasmon_dispersion}
    \omega_{\rm pl}(q) = \sqrt{\omega_{01}^2 + \frac{4\pi e^2}{\bar{\varepsilon}\hbar\omega_{01}} (\alpha_{K}+\alpha_{K^\prime}) q}~,
\end{equation}
where
\begin{equation}\label{eqn:inter-band_velocity}
    \alpha_\xi \equiv \sum_{\lambda\neq\lambda^\prime\in\{0,1\}} (-1)^{\lambda^\prime}\int \frac{d^2\bm{k}}{(2\pi)^2}  f_{\bm{k},\lambda} |{}_\xi\langle \bm k, \lambda|\frac{\hat{p}_x}{m^*}|\bm k, \lambda^\prime\rangle_\xi|^2~.
\end{equation}
Eq.~(\ref{eqn:plasmon_dispersion}) has been obtained by retaining only two valence bands and assuming $\epsilon_{\bm k,0}- \epsilon_{\bm k,1} \approx \hbar\omega_{01}$. Indeed, $\lambda,\lambda^\prime = 0,1$ in Eq.~\eqref{eqn:inter-band_velocity} refer to the first and second valence bands, respectively. As we can see in Fig.~\ref{fig:fig1}(b), the inter-band plasmon dispersion~\eqref{eqn:plasmon_dispersion}---represented by a dotted blue line---properly captures the long-wavelength behavior of the numerically calculated dispersion and, in particular, the value of the plasmon gap at ${\bm q} = {\bm 0}$, i.e.~ $\omega_{\rm pl}(0)=\omega_{01}$.

{\color{blue}{\it Mapping onto a Landau-level problem}.}---Figures~\ref{fig:fig2}(b) and (c) show that the first and second valence bands are rather flat and separated by a gap, akin to two adjacent Landau levels. This energy structure suggests the use of techniques that have been developed in the context of the theory of the quantum Hall effect~\cite{GiulianiVignale}. In these cases, indeed, it is often convenient to project the full Hamiltonian onto a restricted Hilbert space comprising one or two  Landau levels~\cite{girvin_PRB_1984,girvin_PRB_1986}.

For twisted ${\rm MoTe}_2$, the mapping onto a Landau-level model was first proposed in Ref.~\cite{morales-duran_PRL_2024}. The first step in this approach is to project the Hamiltonian~\eqref{eqn:valence_hamiltonian2} onto a single layer-pseudospin sector~\cite{volovik87, bruno_PRL_2004, paul23, morales-duran_PRL_2024,shi_PRB_2024}. This projection is justified when relative energy splitting between the two pseudospin sectors is large, i.e.~when the condition $p^2_0/m^* \ll \Delta$, where $p_0$ is the typical electron momentum and $\Delta$ is the typical value of $|\bm{\Delta}(\bm{r})|$. Alternatively, one can write this condition as $\hbar^2/(m^* A_{\rm M}) \ll \Delta$, where $A_{\rm M}$ is the area of the moir\'e unit cell.

In Ref.~\cite{SM}, we discuss a formalism based on semiclassical techniques~\cite{Belov06,Littlejohn91,Reijnders18}, with which one can perform such projection to any desired order in the dimensionless parameter $\zeta \equiv \hbar^2/(m^* A_{\rm M} \Delta)$.  This approach is rooted into the relation between quantum mechanical operators on Hilbert spaces and classical observables on phase space~\cite{Martinez02,Zworski12,Hall13}. Its main advantage over previously considered methods~\cite{volovik87,bruno_PRL_2004,paul23,morales-duran_PRL_2024,shi_PRB_2024} is that it provides a {\it systematic} way to construct the projection as an expansion in powers of $\zeta$, including all higher-order terms. 

In Ref.~\cite{SM} we show that, to lowest order in the parameter $\zeta$, our effective scalar Hamiltonian for the upper layer-pseudospin sector coincides with the previously obtained result~\cite{volovik87,bruno_PRL_2004,paul23,morales-duran_PRL_2024,shi_PRB_2024}, i.e.
\begin{equation}  \label{eqn:Hamiltonian-effective-single-band}
  \hat{H} = -\frac{1}{2m^*}\left[\hat{\bm{p}} + \frac{e}{c} \bm{A}(\bm{r})\right]^2 - D(\bm{r}) + 
  \Delta_0(\bm{r}) + | \bm{\Delta}(\bm{r}) |~,
\end{equation}
where $\hat{\bm{p}}=-i \hbar \partial_{\bm{r}}$ is the usual momentum operator. The vector potential $\bm{A}$ in Eq.~(\ref{eqn:Hamiltonian-effective-single-band}) is given by
\begin{equation}  \label{eqn:vector-potential-plus}
  \bm{A}(\bm{r}) = \frac{\hbar c}{2e} \frac{1}{1 - n_z} \left( -n_x \frac{\partial n_y}{\partial \bm{r}} + n_y \frac{\partial n_x}{\partial \bm{r}}\right)
\end{equation}
while
\begin{equation}\label{eq:Dofr}
D(\bm{r}) \equiv \frac{\hbar^2}{8 m^*} \left[ (\partial_x \bm{n})^2 + (\partial_y \bm{n})^2 \right]~.
\end{equation}
In Eqs.~(\ref{eqn:vector-potential-plus})-(\ref{eq:Dofr}), $\bm{n}(\bm{r}) \equiv  \bm{\Delta}(\bm{r})/|\bm{\Delta}(\bm{r})|$ is the so-called {\it Skyrme texture}. We remind the reader that the quantities ${\bm \Delta}({\bm r})$ and $\Delta_0({\bm r})$ have been introduced above in Eq.~(\ref{eqn:delta}) and~(\ref{eqn:delta_0}), respectively.  

Since the Skyrme texture has topological charge~\cite{wu_PRL_2019}
\begin{equation}
    N = \frac{1}{4\pi}\int_{A_{\rm M}}  \bm{n}(\bm{r}) \cdot \left[\partial_x \bm{n}(\bm{r}) \cross \partial_y \bm{n}(\bm{r})\right]d^2{\bm r}=-1~,
\end{equation}
the induced real-space magnetic field~\cite{bruno_PRL_2004,morales-duran_PRL_2024} 
\begin{equation}  \label{eqn:magnetic-field-z}
  B_z^{\rm eff}(\bm r) = \frac{\hbar c}{2 e} \bm{n}(\bm{r}) \cdot \left[\partial_x \bm{n}(\bm{r}) \cross \partial_y \bm{n}(\bm{r})\right]
\end{equation}
carries a total magnetic flux $\Phi_0=-hc/e$ through the unit cell of the moir\'e lattice of area $A_{\rm M}$.

As discussed in Ref.~\cite{morales-duran_PRL_2024}, the magnetic field~(\ref{eqn:magnetic-field-z}) has a non-zero average given by $B_{0z}^\mathrm{eff} \equiv \Phi_0/A_{\rm M}$. We can therefore split the vector potential $\bm{A}$ into a linear part $\bm{A}_0$, which gives rise to this constant magnetic field, and a remainder with a zero-average over the moir\'e unit cell.
One subsequently introduces $\hat{\bm{\Pi}} = \hat{\bm{p}} + e \bm{A}_0/c$ and the raising and lowering operators~\cite{morales-duran_PRL_2024,GiulianiVignale}
\begin{equation}  \label{eqn:def-ladder-operators}
  \hat{a}^\dagger = \frac{\ell}{\sqrt{2}\hbar} \left( \hat{\Pi}_y - i \hat{\Pi}_x \right)~, \quad
  \hat{a} = \frac{\ell}{\sqrt{2}\hbar} \left( \hat{\Pi}_y + i \hat{\Pi}_x \right)~,
\end{equation}
where $\ell$ is the effective magnetic length, defined by the condition $2\pi\ell^2 B_{0z}^\mathrm{eff} = \Phi_0$. Using the definition of $B_{0z}^\mathrm{eff}$, we can also write this relation as $2\pi \ell^2 = A_{\rm M}$.

In terms of the ladder operators~\eqref{eqn:def-ladder-operators}, we can write the Hamiltonian~\eqref{eqn:Hamiltonian-effective-single-band} as~\cite{morales-duran_PRL_2024}
\begin{equation}
\label{eqn:H-Landau-levels}
  \hat{H} = - \hbar \omega_{\rm c} \left( \hat{a}^\dagger \hat{a} + \frac{1}{2} \right) + \hat{H}_{\rm cor}~,
\end{equation}
where we have introduced the effective cyclotron frequency
\begin{equation}\label{eq:cyclotron_frequency_1}
\omega_{\rm c} = \frac{e B_{0z}^{\rm eff}}{m^* c}   
\end{equation}
and $\hat{H}_{\rm cor}$ is a ``correction'' term, which was explicitly computed in Ref.~\cite{morales-duran_PRL_2024}.

In absence of $\hat{H}_{\rm cor}$, the Hamiltonian~\eqref{eqn:H-Landau-levels} is simply the Landau-level Hamiltonian. Its spectrum corresponds to perfectly flat bands located at the energies $E_n = - ( n + \tfrac{1}{2} ) \hbar \omega_{\rm c}$. 
In the vicinity of the magic angle, the corrections arising from $\hat{H}_{\rm cor}$ can be neglected for the first valence band ($n=0$), as shown explicitly in Ref.~\cite{morales-duran_PRL_2024}. Figure~\ref{fig:fig2} shows that the second valence band ($n=1$) is also rather flat (with a bandwidth smaller than $10~{\rm meV}$) at $\theta=3.1^\circ$, allowing us to neglect the corrections for this band as well. We therefore conclude that the original Hamiltonian~\eqref{eqn:Hamiltonian-effective-single-band} can be approximately mapped onto a Landau-level Hamiltonian, at least when the aim is to discuss physics involved with the first and second valence bands.

The mapping onto a Landau-level model allows us to analytically compute the contribution to the inter-band optical conductivity due to transitions between the first and second valence bands.
In Ref.~\cite{SM}, we explicitly perform this calculation, starting from an expression similar to Eq.~\eqref{eqn:conductivity_inter-band}. Introducing the Landau-level eigenstates, we obtain
\begin{equation}  \label{eqn:conductivity-Landau-level-model}
  \sigma_{\alpha\beta}^{\rm inter}(\omega) =
    \frac{n_0 e^2}{m^*} \frac{1}{\omega_{\rm c}^2 - (\omega + i\eta)^2} 
    \begin{pmatrix}
      -i\omega & \omega_{\rm c} \\
      -\omega_{\rm c} & -i\omega
    \end{pmatrix}~,
\end{equation}
where $\eta=0^+$. The Landau-level model therefore qualitatively explains the existence of a non-zero (single-spin and single-valley) Hall conductivity $\sigma_{xy}^{\rm inter}(\omega)$ as seen in Fig.~\ref{fig:fig3}(b). In this context,  $\sigma_{xy}^{\rm inter}(\omega)$ arises from the effective magnetic field $B_{0z}^\mathrm{eff}$ in the Hamiltonian~\eqref{eqn:Hamiltonian-effective-single-band}.

We now turn to discuss how gapped plasmon modes emerge in this Landau-level picture. We can calculate the long-wavelength dispersion of inter-band plasmons from the diagonal elements of the conductivity~\eqref{eqn:conductivity-Landau-level-model}. 
Imposing that the real part of the dielectric function~\eqref{eqn:dielectric_tensor} vanishes, we obtain
\begin{equation}
{\rm Re}\left[\varepsilon(\bm{q} \to {\bm 0}
, \omega)\right]_{\bm{0},\bm{0}} = \Re\left[1 + i\frac{2\pi}{q\bar{\varepsilon}} \frac{q^2\sigma_{\rm L}(\omega)}{\omega}\right] = 0~,
\end{equation}
where $\sigma_{\rm L}(\omega) = \sigma_{xx}^{\rm inter}(\omega) = \sigma_{yy}^{\rm inter}(\omega)$. Note that the Hall conductivity does not appear in the plasmon dispersion, because of the usual cancellation of the two valley-resolved contributions in the case that time-reversal symmetry is not broken. Using our result~\eqref{eqn:conductivity-Landau-level-model} for the conductivity, we find that the long-wavelength dispersion of collective modes in our system is given by the famous magnetoplasmon formula~\cite{fetter_1985}
\begin{equation}\label{eqn:magnetoplasmon}
  \omega_{\rm pl}(q) = \sqrt{\omega_{\rm c}^2  + \omega_{\rm 2D}^{2}(q) }~,
\end{equation}
where $\omega_{\rm 2D}(q) = \sqrt{2\pi n_0 e^2 q/(m^* \bar{\varepsilon})}$ is the usual 2D gapless plasmon dispersion in a parabolic-band electron liquid in zero external magnetic field~\cite{GiulianiVignale}. Equation~\eqref{eqn:magnetoplasmon} only gives us the long-wavelength dispersion of the plasmon, since it stems from a local approximation to the optical conductivity tensor. 
Given the Landau-level-type Hamiltonian (\ref{eqn:H-Landau-levels}) one can go beyond the local limit, in principle,  by computing the full non-interacting (bubble or) polarization function $\chi_0(\bm{q},\omega)$~\cite{GiulianiVignale}.
This calculation is performed explicitly in Ref.~\cite{SM}, where we show that it yields the same result as in Eq.~\eqref{eqn:magnetoplasmon} for the plasmon dispersion in the long-wavelength limit.

The analytical results~\eqref{eqn:conductivity-Landau-level-model} for the conductivity and~\eqref{eqn:magnetoplasmon} for the plasmon dispersion are textbook results~\cite{GiulianiVignale} for a 2D electron liquid subject to a homogeneous perpendicular magnetic field.
Their main merit is to qualitatively explain the numerical results presented above.
More precisely, Eq.~\eqref{eqn:conductivity-Landau-level-model} shows that the (single spin and single valley) Hall conductivity $\sigma_{xy}^{\rm inter}(\omega)$ is non-zero, in accordance with Fig.~\ref{fig:fig3}(b). Equation~\eqref{eqn:magnetoplasmon} predicts a gapped plasmon mode, in accordance with Fig.~\ref{fig:fig1}.

However, a quantitative comparison reveals that the predictive power of the Landau-level model is quite limited. Let us first take a closer look at the gap in the plasmon dispersion. According to Eq.~(\ref{eqn:magnetoplasmon}), in the Landau-level-mapping setting discussed in this Section, the plasmon gap at ${\bm q} = {\bm 0}$ coincides with the cyclotron frequency, i.e.~$\hbar \omega_{\rm pl}(0)= \hbar\omega_{\rm c}$. Using the  definition of $B_{0z}^{\rm eff}$ introduced above, i.e.~$B_{0z}^{\rm eff} = \Phi_0/A_{\rm M}$, we can estimate the cyclotron frequency (\ref{eq:cyclotron_frequency_1}) for small twist angles $\theta$ as following:
\begin{equation}\label{eq:cyclotron_frequency_2}
  \hbar \omega_{\rm c} \approx \frac{2\pi\hbar^2}{m^*}\frac{\theta^2}{a^2}~,
\end{equation}
Using the typical values for the parameters $m^*$ and $a$ given above (for ${\rm MoTe}_2$~\cite{wang_prl_2024}), we find $\hbar \omega_{\rm c} \approx 1.94~\theta^2~{\rm meV}$, with $\theta$ expressed in degrees. For $\theta = 3.1^\circ$, this implies  $\hbar\omega_{\rm c} \approx 19~{\rm meV}$.
Comparing this value to the plasmon gap obtained analytically above in Eq.~(\ref{eqn:plasmon_dispersion}), i.e.~$\hbar \omega_{\rm pl}(0)=\hbar\omega_{01} \approx 8~{\rm meV}$, we immediately see that they disagree by more than a factor of $2$. The gap in the plasmon dispersion predicted by the Landau level model, i.e.~Eq.~\eqref{eqn:magnetoplasmon}, is far too large in comparison with the analytical gap (\ref{eqn:plasmon_dispersion}) calculated above, which compares well with the full numerical results shown in Fig.~\ref{fig:fig1}.

The Landau-level model has another, more qualitative difficulty. The second term under the square-root sign in Eq.~(\ref{eqn:magnetoplasmon}) is linear in the wave number $q$ as in the case of the analytical calculation (\ref{eqn:plasmon_dispersion}). The coefficients controlling these linear terms are however dramatically different.  In the Landau-level model, this coefficient is determined by the f-sum rule~\cite{GiulianiVignale}. In Eq.~\eqref{eqn:plasmon_dispersion}, it is instead related to the inter-band Berry connection~\cite{SM,chakraborty_prb_2022}. Eq.~\eqref{eqn:plasmon_dispersion} also takes into account the fact that the bands are not exactly flat.

{\color{blue} {\it Summary and conclusions.}}---We have calculated the optical properties and plasmonic spectrum of twisted homobilayer TMDs, focusing, for purely illustrative properties, on twisted ${\rm MoTe}_2$. Our main numerical results are shown in Fig.~\ref{fig:fig1} and~\ref{fig:fig3}. Results in Fig.~\ref{fig:fig1} are amenable to experimental testing via scattering-type near-field optical spectroscopy~\cite{Basov_Nanophotonics_2021,Low_NatureMaterials_2017,Lundeberg_Science_2017,Woessner_ACS_2017,AlcarazIranzo_Science_2018,ni_nature_2018}. We have then gone one step further trying to understand the link between the real-space topological properties of the skyrmion Chern-band models introduced in Refs.~\cite{morales-duran_PRL_2024,reddy_arxiv_2024} and our numerical predictions. With the caveats discussed in the previous section, we have shown that the bulk plasmon dispersion in the long-wavelength limit roughly probes the uniform, i.e.~spatially-constant, part of the skyrmion magnetic field (\ref{eqn:magnetic-field-z}). Real-space variations of the latter, which are encoded in the correction term $\hat{H}_{\rm cor}$ in Eq.~(\ref{eqn:H-Landau-levels}), are expected to appear in the non-local corrections to the plasmon dispersion relation at finite $q$. In this work we have included electron-electron interactions in the calculation of collective modes at the level of the RPA~\cite{GiulianiVignale}, which well describes weakly correlated electron systems in the long-wavelength limit. In systems with flat bands and strong electron-electron interactions, however, one may want to calculate collective modes with non-perturbative approaches. One of these is certainly brute-force exact diagonalization~\cite{hu_arxiv_2023}. Another one is the Bijl-Feynmann single-mode approximation (SMA), which was first used~\cite{feynman_1954} in the calculation of the phonon-roton spectrum of $^{4}{\rm He}$. This approach was successfully generalized to the case of strongly correlated fractional quantum Hall fluids by Girvin, MacDonald, and Platzman (GMP)~\cite{girvin_PRB_1986,girvin_PRL_1985} leading to the prediction of gapped magneto-rotons in the lowest Landau level. The GMP-SMA has also been applied to the present problem~\cite{wolf_arxiv_2024}. We emphasize, however, that a qualitative difference between fractional quantum Hall fluids and interacting Chern insulators in twisted MoTe$_2$ needs to be taken into account. In the standard GMP theory~\cite{girvin_PRB_1986,girvin_PRL_1985}, Landau-level mixing can indeed be safely neglected. In the present problem, however, Landau-level mixing is not negligible. In fact, the electron-electron interaction energy scale $V_{\rm ee}$ greatly exceeds the effective cyclotron energy. Using indeed the length scale $\sqrt{A_{\rm M}}$ as a typical value of the inter-electron distance, we find:  
\begin{equation}
\frac{V_{\rm ee}}{\hbar\omega_{\rm c}} \sim \frac{\sqrt{A_{\rm M}}}{a_{\rm B}^*} \approx \frac{65}{\bar{\varepsilon}} ~,
\end{equation}
where $a_{\rm B}^* = \bar{\varepsilon}\hbar^2/(m^* e^2)$ is the material's Bohr radius.
The last approximation holds for the choice of parameters we made throughout this work,~i.e. $a = 0.352~{\rm nm}$, $m^* = 0.6~m_{\rm e}$, and $\theta = 3.1^\circ$.

{\color{blue} {\it Acknowledgments.}}---L.C. and M.P. were supported by the MUR - Italian Ministry of University and Research under the ``Research projects of relevant national interest  - PRIN 2020''  - Project No.~2020JLZ52N (``Light-matter interactions and the collective behavior of quantum 2D materials, q-LIMA'') and by the European Union under grant agreement No. 101131579 - Exqiral and No.~873028 - Hydrotronics.  
The work of K.J.A.R and M.I.K. was supported by the European Union's Horizon 2020 research and innovation programme under European Research Council synergy grant 854843 ``FASTCORR''. Views and opinions expressed are however those of the author(s) only and do not necessarily reflect those of the European Union or the European Commission. Neither the European Union nor the granting authority can be held responsible for them.\\
$^*$ These authors contributed equally to this work.\\
$^\dagger$ lorenzo.cavicchi@sns.it

%
\newpage
\setcounter{section}{0}
\setcounter{equation}{0}%
\setcounter{figure}{0}%
\setcounter{table}{0}%

\setcounter{page}{1}

\renewcommand{\thetable}{A\arabic{table}}
\renewcommand{\theequation}{A\arabic{equation}}
\renewcommand{\thefigure}{A\arabic{figure}}
\renewcommand{\theHfigure}{A\arabic{figure}}
\renewcommand{\bibnumfmt}[1]{[A#1]}
\renewcommand{\citenumfont}[1]{A#1}

\onecolumngrid
\newpage
\numberwithin{equation}{section}
%
%
\clearpage 
\newpage

\setcounter{section}{0}
\setcounter{equation}{0}%
\setcounter{figure}{0}%
\setcounter{table}{0}%

\setcounter{page}{1}

\renewcommand{\thetable}{S\arabic{table}}
\renewcommand{\theequation}{S\arabic{equation}}
\renewcommand{\thefigure}{S\arabic{figure}}
\renewcommand{\bibnumfmt}[1]{[S#1]}
\renewcommand{\citenumfont}[1]{S#1}

\onecolumngrid

\begin{center}
\textbf{\Large Supplemental Material for:\\ ``Long-wavelength plasmons as detectors of real space topology in twisted transition metal dichalcogenides homobilayers''}

\bigskip

Lorenzo Cavicchi,$^{1,2}$
Koen J. A. Reijnders,$^{3}$
Mikhail I. Katsnelson,$^{3}$
Marco Polini$^{2,\,4}$

\bigskip

$^1$\!{\it Scuola Normale Superiore, I-56126 Pisa,~Italy}

$^2$\!{\it Dipartimento di Fisica dell'Universit\`a di Pisa, Largo Bruno Pontecorvo 3, I-56127 Pisa,~Italy}

$^3$\!{\it Radboud University, Institute for Molecules and Materials,
Heyendaalseweg 135, 6525 AJ Nijmegen,~The Netherlands}

$^4$\!{\it ICFO-Institut de Ci\`{e}ncies Fot\`{o}niques, The Barcelona Institute of Science and Technology, Av. Carl Friedrich Gauss 3, 08860 Castelldefels (Barcelona),~Spain}

\bigskip

In this Supplemental Material we present a full quantum mechanical derivation of the plasmon dispersion and expand on the relation between the single-valley and single-spin local Hall conductivity observed in twisted MoTe$_2$ and the orbital magnetic moment. Furthermore, we present a rigorous semiclassical approach to the mapping onto a single Landau level and discuss the conductivity and polarization functions in such projected Landau-level model.
 
\end{center}

\onecolumngrid

\appendix
\section*{Section I: Quantum derivation of the gapped plasmon dispersion}

Starting from the inter-band conductivity expression in Eq.~\eqref{eqn:conductivity_inter-band} of the main text, we derive the dispersion of the lowest-energy inter-band plasmon. Since the physical Hall conductivity is zero (when time-reversal symmetry is not spontaneously or externally broken),~i.e. $\sigma_{xy}(\omega) = 0$, this contribution does not enter the plasmon dispersion and we can focus on the longitudinal part only:
\begin{align}\label{eqn:SM_conductivity_inter-band}
    \sigma_{xx}^{\rm inter}(\omega)&=\frac{ie^2}{\hbar} \sum_{\xi = K,K^\prime}\sum_{\lambda\neq\lambda^\prime} \int \frac{d^2\bm{k}}{(2\pi)^2}\left[-\frac{f_{\bm k, \lambda} - f_{\bm k, \lambda^\prime}}{\epsilon_{\bm k,\lambda}- \epsilon_{\bm k,\lambda^\prime}}\right]\frac{{}_\xi\langle \bm k, \lambda|\frac{\hbar}{m^*}\hat{p}_x|\bm k, \lambda^\prime\rangle_\xi \langle \bm k, \lambda^\prime| \frac{\hbar}{m^*}\hat{p}_x|\bm k, \lambda\rangle_\xi}{\epsilon_{\bm k,\lambda}- \epsilon_{\bm k,\lambda^\prime} + \hbar \omega + i \eta} \nonumber \\ 
    &= -\frac{ie^2}{\hbar} \sum_{\xi = K,K^\prime}\sum_{\lambda\neq\lambda^\prime} \int \frac{d^2\bm{k}}{(2\pi)^2}\left[\frac{f_{\bm k, \lambda}|{}_\xi\langle \bm k, \lambda|\frac{\hbar}{m^*}\hat{p}_x|\bm k, \lambda^\prime\rangle_\xi|^2}{\epsilon_{\bm k,\lambda}- \epsilon_{\bm k,\lambda^\prime}}\right]\left[\frac{1}{\epsilon_{\bm k,\lambda}- \epsilon_{\bm k,\lambda^\prime} + \hbar \omega + i \eta} - \frac{1}{\epsilon_{\bm k,\lambda}- \epsilon_{\bm k,\lambda^\prime} - \hbar \omega - i \eta}\right] \nonumber \\
    &= \frac{ie^2}{\hbar} \sum_{\xi = K,K^\prime}\sum_{\lambda\neq\lambda^\prime} \int \frac{d^2\bm{k}}{(2\pi)^2}\left[\frac{f_{\bm k, \lambda}|{}_\xi\langle \bm k, \lambda|\frac{\hbar}{m^*}\hat{p}_x|\bm k, \lambda^\prime\rangle_\xi|^2}{\epsilon_{\bm k,\lambda}- \epsilon_{\bm k,\lambda^\prime}}\right]\left[\frac{2\hbar\omega}{\left(\epsilon_{\bm k,\lambda}- \epsilon_{\bm k,\lambda^\prime}\right)^2 - \left(\hbar \omega + i \eta\right)^2}\right]~.
\end{align}
Since we are interested in the transition between the first and second valence bands, we can focus only on $\lambda, \lambda^\prime=0,1$. Also, since these bands are rather flat, we approximate the transition energy as $\epsilon_{\bm k,\lambda=0} - \epsilon_{\bm k,\lambda^\prime=1}\approx \hbar\omega_{01}>0$, where $\hbar\omega_{01}$ is the transition energy evaluated at the ${\bm k}$ point in the moir\'e BZ at which each band has a saddle point---see Fig.~\ref{fig:fig2}(c) in the main text. This rough approximation captures the frequency behaviour of the optical conductivity around $\omega \approx \omega_{01}$, also in our case of non-perfectly-flat bands~\cite{SM_chakraborty_prb_2022}. Introducing also the inter-band mean value of the longitudinal momentum operator, 
\begin{equation}\label{eqn:alpha}
    \hbar^2 \alpha_\xi \equiv \sum_{\lambda\neq\lambda^\prime\in\{0,1\}} (-1)^{\lambda^\prime}\int \frac{d^2\bm{k}}{(2\pi)^2}  f_{\bm{k},\lambda} |{}_\xi\langle \bm k, \lambda|\frac{\hbar}{m^*}\hat{p}_x|\bm k, \lambda^\prime\rangle_\xi|^2~,
\end{equation}
we find the following result:
\begin{equation}\label{eqn:conductivity_SM}
    \sigma_{xx}^{\rm inter}(\omega) \approx -\frac{2ie^2}{\hbar} \hbar^2(\alpha_K + \alpha_{K^\prime})\frac{\omega}{\omega_{01}}\frac{1}{\hbar^2\omega_{01}^2 - \left(\hbar \omega + i \eta\right)^2}~.
\end{equation}
Substituting expression~\eqref{eqn:conductivity_SM} into the equation for the dielectric function (i.e.~Eq.~\eqref{eqn:dielectric_tensor} of the main text), we get the following plasmon dispersion in the long-wavelength $q\to0$ limit
\begin{equation}\label{eqn:effective_magneto_plasmon}
    \omega_{\rm pl}(q) = \sqrt{\omega_{01}^2 + \frac{4\pi e^2}{\bar{\varepsilon}\hbar\omega_{01}} (\alpha_K + \alpha_{K^\prime}) q}~,
\end{equation}
which coincides with Eq.~(\ref{eqn:plasmon_dispersion}) of the main text. This result is represented in Fig.~\ref{fig:fig1}(b) of the main text as a dotted blue line.

We now demonstrate that the quantity $\alpha_\xi$ introduced in Eq.~\eqref{eqn:alpha} is related to the Berry inter-band connection~\cite{SM_chakraborty_prb_2022}. To this aim, the following relation is useful:
\begin{equation}\label{eqn:momentum_matrix_element_identity}
    {}_\xi\langle \bm k, \lambda|\frac{\hbar}{m^*}\hat{\bm p}|\bm k, \lambda^\prime\rangle_\xi = {}_\xi\langle u_{\bm{k}\lambda}|\partial_{\bm k} \hat{\cal H}(\bm{k}) |u_{\bm{k}\lambda^\prime}\rangle_\xi = {}_\xi\langle \partial_{\bm k}u_{\bm{k}\lambda}|u_{\bm{k}\lambda^\prime}\rangle_\xi(\epsilon_{\bm{k}\lambda} - \epsilon_{\bm{k}\lambda^\prime})~,
\end{equation}
where the Bloch state $|\bm{k},\lambda\rangle_\xi$, with eigenvalue $\epsilon_{\bm{k},\lambda}$, has been explicitly written as
\begin{equation}\label{eq:single-particle-Bloch-state}
    |\bm{k},\lambda \rangle_\xi = e^{i\bm{k}\cdot\bm{r}}|u_{\bm{k}\lambda}\rangle_\xi~.
\end{equation}
Using this result in Eq.~\eqref{eqn:alpha}, we obtain:
\begin{align}\label{eqn:alpha_berry}
    \hbar^2 \alpha_\xi &= \sum_{\lambda\neq\lambda^\prime\in\{0,1\}} (-1)^{\lambda^\prime}\int \frac{d^2\bm{k}}{(2\pi)^2}  f_{\bm{k},\lambda} |\bm{x}\cdot{}_\xi\langle \partial_{\bm k}u_{\bm{k}\lambda}|u_{\bm{k}\lambda^\prime}\rangle_\xi(\epsilon_{\bm{k}\lambda} - \epsilon_{\bm{k}\lambda^\prime})|^2 \nonumber\\
    &\approx \hbar^2\omega_{01}^2\sum_{\lambda\neq\lambda^\prime\in\{0,1\}} (-1)^{\lambda^\prime}\int \frac{d^2\bm{k}}{(2\pi)^2}  f_{\bm{k},\lambda} |\bm{x}\cdot\bm{\mathcal{A}}_{\lambda,\lambda^\prime}^\xi(\bm{k})|^2~,
\end{align}
where we have approximated $(\epsilon_{\bm{k}\lambda} - \epsilon_{\bm{k}\lambda^\prime})^2\approx\hbar^2\omega_{01}^2$ and introduced the inter-band Berry connection
\begin{equation}
\bm{\mathcal{A}}_{\lambda,\lambda^\prime}^\xi(\bm{k}) \equiv i{}_\xi\langle \partial_{\bm k} u_{\bm{k}\lambda}|u_{\bm{k}\lambda^\prime}\rangle_\xi~.
\end{equation}
\section*{Section II: Orbital magnetic moment and inter-band single-valley and single-spin Hall conductivity}
\begin{figure}[b]
    \centering
    \begin{tabular}{lll}
        \begin{overpic}[width=0.5\textwidth]{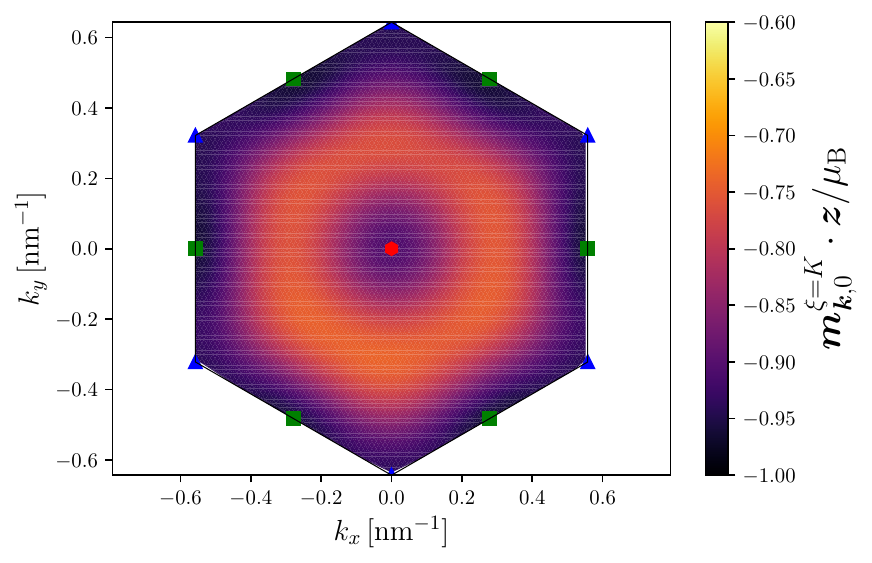}
            \put(0,63){(a)}
        \end{overpic}
        & \begin{overpic}[width=0.5\textwidth]{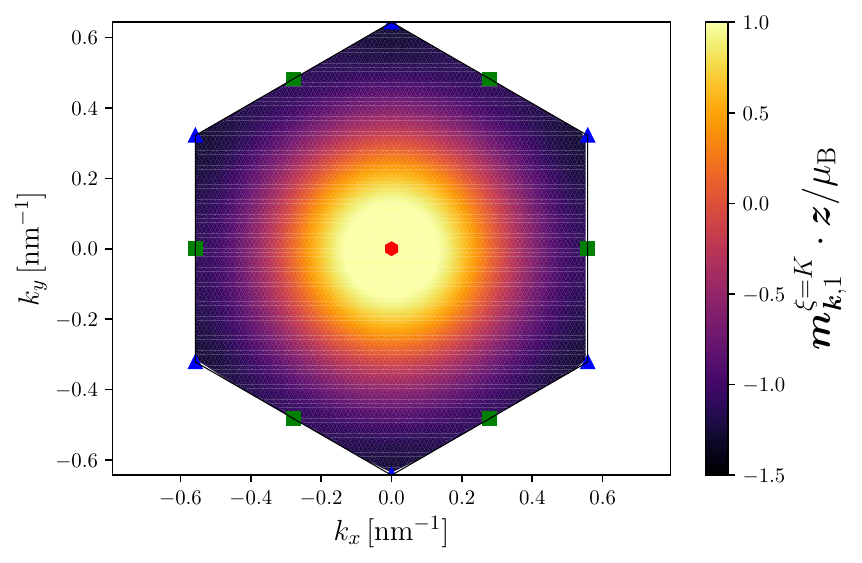}
            \put(0,63){(b)}
        \end{overpic}
    \end{tabular}
    
    \caption{(Color online) Orbital magnetic moment of Bloch states $\bm{m}_{\bm{k},\lambda}^{\xi = K}$ in units of the Bohr magneton $\mu_{\rm B} = e\hbar/(2m_{\rm e} c)$ for twisted ${\rm MoTe}_2$ at a twist angle $\theta = 3.1^\circ$. All the other microscopic parameters are the same as in Fig.~\ref{fig:fig2} of the main text. Results in this figure have been obtained from Eq.~\eqref{eqn:bloch_magnetization}. Panel a) [Panel b)]: results for the first $\lambda = 0$ [second $\lambda = 1$] valence band. The $\bm k$-dependence is shown throughout the moir\'e BZ. In both panels, small colored symbols refer to high symmetry points: the red hexagon is the $\Gamma$ point; the green squares are the $M$/$M^\prime$ points; the blue triangles are the $K$/$K^\prime$ points.}
    \label{fig:SM1}
\end{figure}
\begin{figure}
    \centering
    \begin{tabular}{lll}
        \begin{overpic}[width=0.5\textwidth]{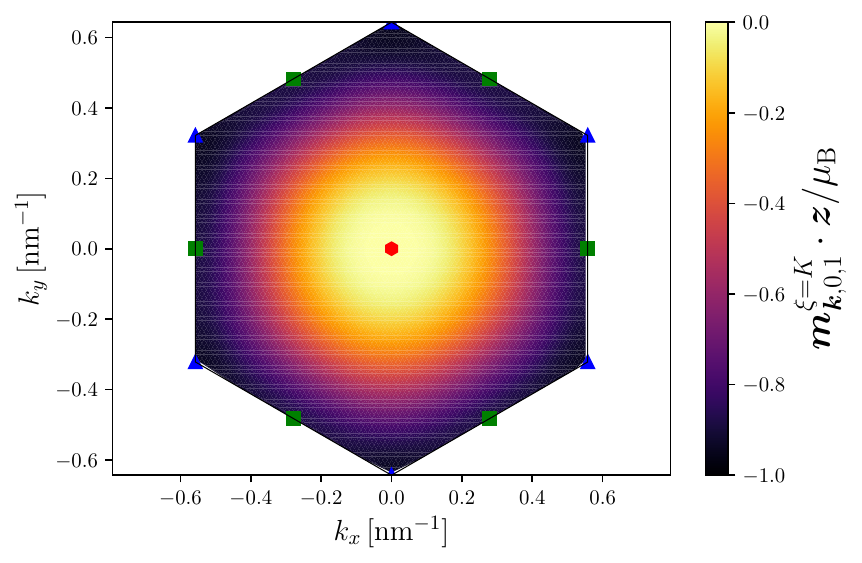}
            \put(0,63){(a)}
        \end{overpic}
        & \begin{overpic}[width=0.5\textwidth]{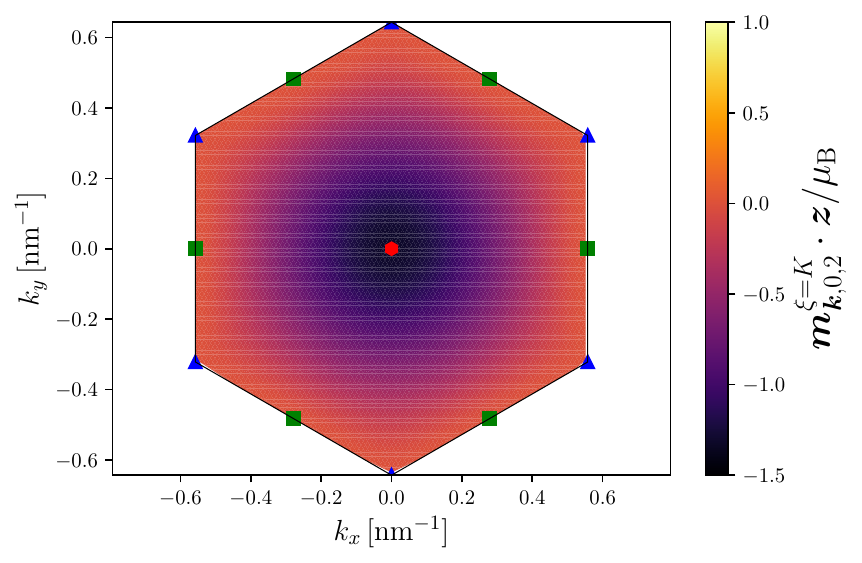}
            \put(0,63){(b)}
        \end{overpic}
    \end{tabular}
    
    \caption{(Color online) Numerical results for the inter-band orbital magnetic moment of Bloch states $\bm{m}_{\bm k,\lambda,\lambda^\prime}^{\xi = K}$ in units of the Bohr magneton $\mu_{\rm B} = e\hbar/(2m_{\rm e} c)$. This quantity appears in the formula for the Hall component of the conductivity tensor, Eq.~\eqref{eqn:sigma_xy_orbital}, and its explicit expression has been reported in Eq.~\eqref{eqn:bloch_orbital_momentum} of the main text. Panel (a) shows results for $\lambda =0 $ and $\lambda^\prime =1 $; panel (b) shows results for $\lambda =0 $ and $\lambda^\prime =2$. The twist angle is fixed at $\theta = 3.1^\circ$ while all the other microscopic parameters are the same as in Fig.~\ref{fig:fig2} of the main text. The $\bm k$-dependence is shown throughout the moir\'e BZ. In both panels, small colored symbols refer to high symmetry points: the red hexagon is the $\Gamma$ point; the green squares are the $M$/$M^\prime$ points; the blue triangles are the $K$/$K^\prime$ points.}
    \label{fig:SM2}
\end{figure}
The orbital magnetic moment $\bm{m}_{\bm{k}\lambda}^\xi$ of a Bloch state $|\bm k, \lambda\rangle_\xi$ is given by~\cite{SM_vanderbilt}:
\begin{equation}\label{eqn:bloch_magnetization}
    \bm{m}_{\bm{k},\lambda}^\xi = \frac{-ie}{2\hbar c}{}_\xi\langle\partial_{\bm{k}}u_{\bm{k},\lambda}|\times[\hat{\cal H}(\bm{k}) - \epsilon_{\bm{k},\lambda}]|\partial_{\bm{k}}u_{\bm{k},\lambda}\rangle_\xi~,
\end{equation}
where $\hat{\cal H}(\bm{k}) \equiv e^{-i\bm{k}\cdot\bm{r}} \hat{\cal H} e^{i\bm{k}\cdot\bm{r}}$. The expression given in Eq.~\eqref{eqn:bloch_magnetization} for the orbital magnetic moment of Bloch electrons was first proposed in Refs.~\cite{SM_niu_1996, SM_niu_1999} and stems form a semi-classical description of electrons roaming in a crystal. 

Eq.~(\ref{eqn:bloch_magnetization}) can be further handled using the second equality in Eq.~\eqref{eqn:momentum_matrix_element_identity}, obtaining:
\begin{equation}\label{eqn:bloch_orbital_momentum_2}
    {\bm m}_{{\bm k},\lambda}^\xi = \frac{ie}{2\hbar c}\sum_{\lambda^\prime}\left[\frac{{}_\xi\langle u_{\bm{k},\lambda}|\partial_{\bm k} \hat{\cal H}(\bm{k}) |u_{\bm{k},\lambda^\prime}\rangle_\xi\times{}_\xi\langle u_{\bm{k},\lambda^\prime}|\partial_{\bm k} \hat{\cal H}(\bm{k}) |u_{\bm{k},\lambda}\rangle_\xi}{\epsilon_{\bm{k},\lambda} - \epsilon_{\bm{k},\lambda^\prime}}\right]~.
\end{equation}
This expression has a clear advantage with respect to the previous one in that no derivatives acting on states appear in it and it is manifestly gauge invariant.

When time-reversal symmetry is preserved, $\bm{m}_{\bm{k},\lambda}^{\xi} = -\bm{m}_{-\bm{k},\lambda}^{\bar{\xi}}$. If inversion symmetry is also preserved, we get  $\bm{m}_{\bm{k},\lambda}^\xi = \bm{m}_{-\bm{k},\lambda}^{\bar{\xi}}$, yielding $\bm{m}_{\bm{k},\lambda}^{\xi}=0$, $\forall \lambda$. Notice that the application of both time-reversal (${\cal T}$) and inversion ($C_2$) flips the valley index $\xi\to\bar{\xi}$, where $\bar{\xi} = K^\prime,K$ when $\xi = K,K^\prime$, respectively. At equilibrium, the total orbital magnetization is zero due to the usual cancellation between valley-resolved contributions. However, since we are considering a single-valley and single-spin model Hamiltonian with broken $C_2{\cal T}$ symmetry, the orbital magnetization of single-valley Bloch states can be nonzero. In Fig.~\ref{fig:SM1} we show results for $\bm{m}_{\bm{k},\lambda}^{\xi=K}$ as obtained from Eq.~\eqref{eqn:bloch_orbital_momentum_2}, for the case of twisted MoTe$_2$ and $\lambda=0,1$.

We now show that the single-valley and single-spin inter-band Hall conductivity is related to the orbital magnetic moment as defined in Eq.~\eqref{eqn:bloch_orbital_momentum}. We use the first identity in Eq.~\eqref{eqn:momentum_matrix_element_identity} into the expression for the inter-band conductivity, i.e.~Eq.~\eqref{eqn:conductivity_inter-band} in the main text, obtaining:
\begin{align}
\sigma_{xy,\xi}^{\rm inter}(\omega) &= \frac{ie^2}{2\hbar} \sum_{\lambda\neq\lambda^\prime} \int \frac{d^2\bm{k}}{(2\pi)^2}\left[-\frac{f_{\bm k, \lambda} - f_{\bm k, \lambda^\prime}}{\epsilon_{\bm k,\lambda}- \epsilon_{\bm k,\lambda^\prime}}\right]\frac{{}_\xi\langle u_{\bm{k},\lambda}|\partial_{k_x} \hat{\cal H}(\bm{k}) |u_{\bm{k},\lambda^\prime}\rangle_\xi\langle u_{\bm{k},\lambda^\prime}|\partial_{k_y} \hat{\cal H}(\bm{k}) |u_{\bm{k},\lambda}\rangle_\xi - (x\leftrightarrow y)}{\epsilon_{\bm k,\lambda}- \epsilon_{\bm k,\lambda^\prime} + \hbar \omega + i \eta}\nonumber\\
&=\frac{ie^2}{2\hbar} \sum_{\lambda\neq\lambda^\prime} \int \frac{d^2\bm{k}}{(2\pi)^2}\left[-\frac{f_{\bm k, \lambda} - f_{\bm k, \lambda^\prime}}{\epsilon_{\bm k,\lambda}- \epsilon_{\bm k,\lambda^\prime}}\right]\frac{{}_\xi\langle u_{\bm{k},\lambda}|\partial_{\bm k} \hat{\cal H}(\bm{k}) |u_{\bm{k},\lambda^\prime}\rangle_\xi\times{}_\xi\langle u_{\bm{k},\lambda^\prime}|\partial_{\bm k} \hat{\cal H}(\bm{k}) |u_{\bm{k},\lambda}\rangle_\xi}{\epsilon_{\bm k,\lambda}- \epsilon_{\bm k,\lambda^\prime} + \hbar \omega + i \eta}\cdot\bm{z}\nonumber\\
&=- ec \sum_{\lambda\neq\lambda^\prime} \int \frac{d^2\bm{k}}{(2\pi)^2}\frac{f_{\bm k, \lambda} - f_{\bm k, \lambda^\prime}}{\epsilon_{\bm k,\lambda}- \epsilon_{\bm k,\lambda^\prime}+ \hbar \omega + i \eta}\bm{m}_{\bm k,\lambda,\lambda^\prime}^\xi\cdot\bm{z}~,
\end{align}
where $\bm{m}_{\bm k,\lambda,\lambda^\prime}^\xi$, whose explicit expression has been reported in Eq.~(\ref{eqn:bloch_orbital_momentum}) of the main text, is the inter-band orbital magnetic moment, which is related to related to $\bm{m}_{\bm k,\lambda}^\xi$ in Eq.~(\ref{eqn:bloch_orbital_momentum_2}) by $\bm{m}_{\bm k,\lambda}^\xi = \sum_{\lambda^\prime} \bm{m}_{\bm k,\lambda,\lambda^\prime}^\xi$. In Fig.~\ref{fig:SM2} we report numerical results for $\bm{m}_{\bm k,\lambda,\lambda^\prime}^\xi$ for $\lambda \neq \lambda^\prime = 0,1$ (panel (a)) and $\lambda \neq \lambda^\prime = 0,2$ (panel (b)). 

\begin{figure}
    \centering
    \includegraphics[width=0.5\textwidth]{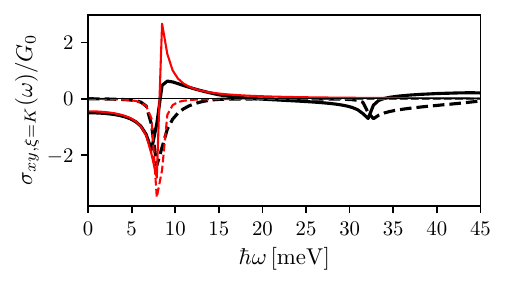}    
    \caption{(Color online) Single-valley and single-spin Hall optical conductivity of twisted ${\rm MoTe}_2$. Black lines refer to the results presented in the main text---see Fig.~\ref{fig:fig3}(b)---obtained from a full Kubo formula calculation based on the valence band Hamiltonian (\ref{eqn:valence_hamiltonian}). Red lines refer to the calculation of the orbital Hall optical conductivity in Eq.~\eqref{eqn:Hall_orbital}. Solid (dashed) lines refer to the real (imaginary) part. The twist angle is fixed to $\theta = 3.1^\circ$.}
    \label{fig:SM3}
\end{figure}

We continue manipulating the expression for $\sigma_{xy,\xi}^{\rm inter}(\omega)$ obtained above:
\begin{align}\label{eqn:sigma_xy_orbital}
    \sigma_{xy,\xi}^{\rm inter}(\omega) &= - ec \sum_{\lambda\neq\lambda^\prime} \int \frac{d^2\bm{k}}{(2\pi)^2}\frac{f_{\bm k, \lambda} - f_{\bm k, \lambda^\prime}}{\epsilon_{\bm k,\lambda}- \epsilon_{\bm k,\lambda^\prime}+ \hbar \omega + i \eta}\bm{m}_{\bm k,\lambda,\lambda^\prime}^\xi\cdot\bm{z}\nonumber\\
    &= - ec \sum_{\lambda\neq\lambda^\prime} \int \frac{d^2\bm{k}}{(2\pi)^2}f_{\bm k,\lambda}\left[\frac{\bm{m}_{\bm k,\lambda,\lambda^\prime}^\xi \cdot\bm{z} }{\epsilon_{\bm k,\lambda}- \epsilon_{\bm k,\lambda^\prime}+ \hbar \omega + i \eta}  - \frac{\bm{m}_{\bm k,\lambda^\prime,\lambda}^\xi \cdot\bm{z}}{\epsilon_{\bm k,\lambda^\prime}- \epsilon_{\bm k,\lambda}+ \hbar \omega + i \eta}\right] \nonumber \\
    &= - 2ec \sum_{\lambda\neq\lambda^\prime} \int \frac{d^2\bm{k}}{(2\pi)^2}f_{\bm k,\lambda} \frac{(\epsilon_{\bm k,\lambda}- \epsilon_{\bm k,\lambda^\prime}) }{(\epsilon_{\bm k,\lambda}- \epsilon_{\bm k,\lambda^\prime})^2 - (\hbar \omega + i\eta)^2}\bm{m}_{\bm k,\lambda,\lambda^\prime}^\xi\cdot\bm{z}~,
\end{align}
where we used that $\bm{m}_{\bm k,\lambda^\prime,\lambda}^\xi = \left(\bm{m}_{\bm k,\lambda,\lambda^\prime}^\xi\right)^*$ and that $\bm{m}_{\bm k,\lambda,\lambda^\prime}^\xi \in\mathbb{R}$. Isolating the contribution from the first inter-band transition occurring between the first and second valence bands, we further get:
\begin{equation}\label{eqn:Hall_orbital}
    \sigma_{xy,\xi}^{\rm inter}(\omega) \approx -G_{\rm mag}^\xi \frac{\hbar\omega_{01}}{\hbar^2\omega_{01}^2 - (\hbar \omega + i\eta)^2}~,
\end{equation}
where, as usual, $\epsilon_{\bm k, 0}- \epsilon_{\bm k, 1} \approx \hbar\omega_{01}$ and we have introduced
\begin{equation}
    G_{\rm mag}^\xi \equiv 2ec \int \frac{d^2\bm{k}}{(2\pi)^2}\Re\left[f_{\bm k, 0}\bm{m}_{\bm k, 0, 1}^\xi - f_{\bm k, 1}\bm{m}_{\bm k, 1, 0}^\xi\right]\cdot\bm{z}~.
\end{equation}
In Fig.~\ref{fig:SM3} we compare the analytical expression~\eqref{eqn:Hall_orbital} with the numerical results presented in Fig.~\ref{fig:fig3}(b) of the main text.

\section*{Section III: Projection onto a single pseudospin sector using semiclassical techniques}

In this Section, we discuss a formalism to extract an effective scalar Hamiltonian, corresponding to a single layer-pseudospin sector, from the matrix Hamiltonian~\eqref{eqn:valence_hamiltonian2}, which comprises both layer-pseudospin sectors. Compared to methods that were previously used to perform this projection~\cite{SM_morales-duran_PRL_2024,SM_bruno_PRL_2004}, our formalism provides a systematic way to construct all terms of higher order in the relative energy splitting $\zeta \equiv \hbar^2/(m^* A_{\rm M} |\bm{\Delta}|)$. 
At the heart of our method, which is an application of the more general formalism discussed in Refs.~\cite{SM_Belov06,SM_Littlejohn91,SM_Reijnders18}, is the relation between quantum mechanical operators on Hilbert space and classical observables on phase space. More precisely, we first pass from quantum mechanical operators to classical observables, and then express the projection onto a scalar Hamiltonian in terms of these classical observables, which are easier to manipulate than their quantum mechanical counterparts. In this way, we obtain an effective classical Hamiltonian, which we subsequently quantize in order to obtain the quantum mechanical Hamiltonian corresponding to a single layer-pseudospin sector.
Since the semiclassical techniques that are required to perform this analysis are not very often discussed in the physics literature, we also provide a brief pedagogical introduction to the necessary mathematical background~\cite{SM_Martinez02,SM_Zworski12,SM_Hall13}.

\subsection*{III.A: Brief review of semiclassical quantization}

In general, the relation between a classical observable and a quantum mechanical operator is not unique. Consider for instance $A(\bm{r},\bm{p}) = \langle \bm{r}, \bm{p} \rangle$, where the angular brackets denote the Cartesian inner product of the vectors $\bm{r}$ and $\bm{p}$. 
Because the quantum mechanical operators $\bm{r}$ and $\hat{\bm{p}}$ do not commute, one can write down three different quantum mechanical operators that correspond to $A(\bm{r},\bm{p})$, namely $\hat{A}^{(0)} = \langle \bm{r}, \hat{\bm{p}} \rangle$, $\hat{A}^{(1)} = \langle \hat{\bm{p}}, \bm{r} \rangle$ and $\hat{A}^{W} = \tfrac{1}{2}(\hat{A}^{(0)} + \hat{A}^{(1)})$. From a physical point of view, one would like real observables to correspond to Hermitian operators, which is only satisfied by $\hat{A}^{W}$.

More generally, one may consider classical observables that are polynomials of $\bm{r}$ and $\bm{p}$. In the so-called Weyl quantization scheme~\cite{SM_Hall13}, the corresponding operator is constructed by averaging over all possible orderings of the operators $\bm{r}$ and $\hat{\bm{p}}$, which gives rise to a Hermitian differential operator.
This quantization procedure has subsequently been extended to functions of $\bm{r}$ and $\bm{p}$ that are, roughly speaking, bounded by polynomials~\cite{SM_Martinez02,SM_Zworski12}. The resulting operators are called pseudodifferential operators, and are not necessarily polynomials of $\hat{\bm{p}}$. Although all of the operators in this Section will be differential operators, we make use of the calculus for pseudodifferential operators, since it allows us to formulate the method in a general way.

In order to precisely formulate the relation between functions $a(\bm{r},\bm{p})$ on classical phase space and quantum operators $\hat{a}$, we specify how they act on a test function $f(\bm{r})$ in the Hilbert space $L^2(\mathbb{R}^d)$, where  $d$ is the dimensionality of space.
Within Weyl quantization, we can write~\cite{SM_Martinez02,SM_Zworski12,SM_Hall13}
\begin{equation} \label{eqn:quantization-Weyl}
  (\hat{a} f)(\bm{r}) = \frac{1}{(2\pi \hbar)^d} \int e^{i \langle \bm{p} , \bm{r}-\bm{r}' \rangle/\hbar} a\left( \frac{\bm{r}+\bm{r}'}{2}, \bm{p}, \hbar\right) f(\bm{r}') \text{d}\bm{r}' \text{d}\bm{p}~.
\end{equation}
The function $a$ is also called the symbol of the operator $\hat{a}$, and is sometimes also denoted by $\sigma(\hat{a})$.
Equation~\eqref{eqn:quantization-Weyl} establishes a one-to-one correspondence between symbols and pseudodifferential operators.

Since Eq.~\eqref{eqn:quantization-Weyl} is rather abstract, we consider a few special cases.
First of all, when a function $a(\bm{r},\bm{p})$ does not contain products of $\bm{r}$ and $\bm{p}$, the corresponding operator $\hat{a}$ is found by simply replacing the variable $\bm{p}$ by the operator $\hat{\bm{p}}$.
The opposite also holds. Hence, the symbol ${\cal H}(\bm{r},\bm{p})$ of the operator $\hat{\cal H}$ in Eq.~\eqref{eqn:valence_hamiltonian2} is given by
\begin{equation}  \label{eqn:valence_hamiltonian2-symbol}
  {\cal H}(\bm{r},\bm{p}) = -\frac{\bm{p}^2}{2 m^*} \tau_0 + \bm{\Delta}(\bm{r}) \cdot \bm{\tau} + \Delta_0(\bm{r}) \tau_0~.
\end{equation}
As a second example~\cite{SM_Hall13}, the symbol $\langle \bm{f}(\bm{r}) , \bm{p} \rangle$ corresponds to the Hermitian operator $\tfrac{1}{2}( \langle \bm{f}(\bm{r}) , \hat{\bm{p}} \rangle + \langle \hat{\bm{p}} , \bm{f}(\bm{r}) \rangle)$.
Finally, we note that when $a$ is a polynomial of $\bm{r}$ and $\bm{p}$, Eq.~\eqref{eqn:quantization-Weyl} gives the same result as averaging over all possible orderings of the operators $\bm{r}$ and $\hat{\bm{p}}$~\cite{SM_Hall13}.

In the cases that we consider in this text, the symbol $a$ has an expansion in powers of $\hbar$, that is,
\begin{equation}  \label{eqn:symbol-expansion-powers}
  a(\bm{r}, \bm{p}, \hbar) = \sum_{j=0}^\infty a_j(\bm{r},\bm{p}) \hbar^j~,
\end{equation}
where the equality should formally be understood as an asymptotic equivalence~\cite{SM_Martinez02}.
The leading-order term $a_0(\bm{r},\bm{p})$ of this expansion is called the principal symbol and can be interpreted as the classical observable on phase space that corresponds to the quantum operator $\hat{a}$. The second term $a_1(\bm{r},\bm{p})$ is called the subprincipal symbol.

In what follows, we will need to determine the symbol of an operator product. Let $\hat{a}$ and $\hat{b}$ therefore be two pseudodifferential operators, with symbols $a$ and $b$, respectively. Then the symbol of their product $\hat{a}\hat{b}$ is given by~\cite{SM_Martinez02,SM_Zworski12}
\begin{equation}  \label{eqn:symbol-operator-product}
  \sigma(\hat{a} \hat{b})(\bm{r},\bm{p},\hbar) = \exp\left[ -\frac{i\hbar}{2} \left( \left\langle \frac{\partial}{\partial \bm{p}'} , \frac{\partial}{\partial \bm{r}} \right\rangle - \left\langle \frac{\partial}{\partial \bm{r}'} , \frac{\partial}{\partial \bm{p}} \right\rangle \right) \right] \Big. a(\bm{r}',\bm{p}',\hbar) b(\bm{r},\bm{p},\hbar) \Big|_{\bm{r}'=\bm{r},\bm{p}'=\bm{p}}~,
\end{equation}
which is also known as the Moyal product of the symbols $a$ and $b$.
We remark that the terms of zeroth order and first order in the product Eq.~(\ref{eqn:symbol-operator-product}) yield
\begin{equation} \label{eqn:symbol-operator-product-leading}
  \sigma(\hat{a} \hat{b}) = a b + \frac{i \hbar}{2} \{ a , b \} + \mathcal{O}(\hbar^2)~,
\end{equation}
where $\{a , b \}$ stands for the Poisson bracket of $a$ and $b$, defined by
\begin{equation}  \label{eqn:Poisson-bracket}
  \{a, b\} = \left\langle \frac{\partial a}{\partial \bm{r}} , \frac{\partial b}{\partial \bm{p}} \right\rangle - \left\langle \frac{\partial a}{\partial \bm{p}} , \frac{\partial b}{\partial \bm{r}} \right\rangle~.
\end{equation}
We note that Eq.~\eqref{eqn:symbol-operator-product-leading} implies that the symbol of the commutator $[\hat{a},\hat{b}]$ is given by $\sigma([\hat{a},\hat{b}]) = i\hbar\{ a , b \} + \mathcal{O}(\hbar^2)$, which agrees with the familiar notion that Poisson brackets in classical mechanics correspond to commutators in quantum mechanics.

\subsection*{III.B: Projection onto a single pseudospin sector: general formalism}

We now review the general formalism to project a matrix Hamiltonian onto a single pseudospin sector. This explanation is based on Refs.~\cite{SM_Belov06,SM_Littlejohn91}, and draws on the theory reviewed in the previous Section. For simplicity, we assume that the original matrix Hamiltonian has two pseudospin sectors, but this is not a fundamental limitation.

Our starting point is the time-independent eigenvalue equation for the Hamiltonian $\hat{\cal H}$
\begin{equation}  \label{eqn:eigenvalue-H}
  \hat{\mathcal{H}} \Psi = E \Psi~,
\end{equation}
where $\Psi$ is a $2\times 1$ pseudospinor.
We would like to compute the pseudospinors $\Psi^\pm$ corresponding to the upper ($+$) and lower ($-$) pseudospin sectors.
To this end, we write down $\Psi^\pm$ as~\cite{SM_Belov06,SM_Littlejohn91}
\begin{equation}  \label{eqn:chi-Ansatz}
  \Psi^\pm = \hat{\chi}^\pm \psi^\pm~,
\end{equation}
where $\psi^\pm$ is now a scalar wavefunction for one of the two pseudospin sectors. The operator $\hat{\chi}^\pm = ( \hat{\chi}^\pm_b , \hat{\chi}^\pm_t )^T$ is, generally speaking, a pseudodifferential operator that takes the form of a $2\times 1$ pseudospinor. From a physical point of view, Eq.~\eqref{eqn:chi-Ansatz} can be viewed as a generalization of the Born-Oppenheimer Ansatz, which uses a function $\chi$ instead of an operator. The operator $\hat{\chi}$ can therefore be interpreted as a generalization of the instantaneous eigenfunction in the Born-Oppenheimer approximation, cf. the extensive discussion in Ref.~\cite{SM_Belov06}. Moreover, the Foldy-Wouthuysen transformation for the Dirac Hamiltonian can also be cast~\cite{SM_Bruening11} in the form of Eq.~\eqref{eqn:chi-Ansatz}.

We now demand that the scalar wavefunction satisfies the eigenvalue equation
\begin{equation}  \label{eqn:eigenvalue-H-eff}
  \hat{H}^\pm \psi^\pm = E \psi^\pm~,
\end{equation}
where $\hat{H}^\pm$ is the scalar Hamiltonian corresponding to the pseudospin sector $\pm$. 
Moreover, we require conservation of probability, in the sense that $\langle \Psi | \Psi \rangle = \langle \psi | \psi \rangle $, where $\langle f | g \rangle$ is the inner product in the appropriate Hilbert space. Through Eq.~(\ref{eqn:chi-Ansatz}), this is equivalent to the requirement that
\begin{equation}  \label{eqn:chi-unitary}
  \hat{\chi}^\dagger \hat{\chi} = 1~,
\end{equation}
where $\hat{\chi}^\dagger$ is the adjoint of $\hat{\chi}$.

Equations~(\ref{eqn:eigenvalue-H})--(\ref{eqn:chi-unitary}) now completely determine the effective Hamiltonians $\hat{H}^\pm$. Combining Eqs.~(\ref{eqn:eigenvalue-H})--(\ref{eqn:eigenvalue-H-eff}), we have
\begin{equation}
  \hat{\mathcal{H}} \hat{\chi}^\pm \psi^\pm = \hat{\mathcal{H}} \Psi = E \Psi 
  = \hat{\chi}^\pm E \psi^\pm = \hat{\chi}^\pm \hat{H}^\pm \psi^\pm~.
\end{equation}
The left-hand side and right-hand side of this equation will certainly be equal when the operator equation
\begin{equation}  \label{eqn:operator-eigenvalue-diag}
  \hat{\mathcal{H}} \hat{\chi}^\pm = \hat{\chi}^\pm \hat{H}^\pm
\end{equation}
is satisfied~\cite{SM_Belov06}. Using the conservation of probability Eq.~(\ref{eqn:chi-unitary}), we can also write this equation as $(\hat{\chi}^\pm)^\dagger \hat{\mathcal{H}} \hat{\chi}^\pm = \hat{H}^\pm$. This shows that we are effectively diagonalizing the operator $\hat{\mathcal{H}}$, where the operators $\hat{\chi}^\pm$ play the role of the eigenvectors and $\hat{H}^\pm$ the role of the eigenvalues. Combining the two vectors into a $2 \times 2$ pseudodifferential operator $\hat{U} = ( \hat{\chi}^+, \hat{\chi}^- )$, we have~\cite{SM_Littlejohn91}
\begin{equation}  \label{eqn:operator-eigenvalue-diag-matrix}
  \hat{U}^\dagger \hat{\mathcal{H}} \hat{U} = \hat{H}~, \qquad \hat{U}^\dagger \hat{U} = \hat{U} \hat{U}^\dagger = 1~,
\end{equation}
where $\hat{H}$ is a diagonal $2 \times 2$ pseudodifferential operator with $\hat{H}^\pm$ on the diagonal.

In the next step of the procedure~\cite{SM_Belov06,SM_Littlejohn91}, one uses the semiclassical techniques discussed in the previous Section. Specifically, one uses Eq.~(\ref{eqn:symbol-operator-product}) to convert Eqs.~\eqref{eqn:operator-eigenvalue-diag} and~\eqref{eqn:chi-unitary} into equations for the symbols corresponding to the individual operators.
One subsequently expands each of these symbols into a power series in $\hbar$, see Eq.~\eqref{eqn:symbol-expansion-powers}.
Gathering all terms of a given order in $\hbar$, one obtains a set of equations that can be solved to obtain the classical symbol $H^\pm$ of the effective scalar Hamiltonian up to the given order in $\hbar$. The last step in the procedure is to use Eq.~\eqref{eqn:quantization-Weyl} to convert this classical symbol into a quantum mechanical Hamiltonian corresponding to a single layer-pseudospin sector. In the next subsection, we perform these different steps for our Hamiltonian~\eqref{eqn:valence_hamiltonian2}.

An important advantage of the formalism discussed in this subsection, is that it provides a systematic way to obtain the projected Hamiltonian up to any desired order in $\hbar$. However, this scalar Hamiltonian will generally be a pseudodifferential operator rather than a differential operator~\cite{SM_Belov06}, and usually contains terms of all orders in $\hbar$, even if the original matrix Hamiltonian does not.
Equation~\eqref{eqn:operator-eigenvalue-diag-matrix} shows that this formalism can also be viewed as a semiclassical version of the Schrieffer-Wolff transformation~\cite{SM_Bravyi11}, in which one uses commutators to diagonalize a matrix Hamiltonian order by order in $\hbar$.

\subsection*{III.C: Projection onto a single pseudospin sector for the Hamiltonian~\eqref{eqn:valence_hamiltonian2}}
Before we perform the projection onto a single layer-pseudospin sector, let us first take a closer look at our expansion parameter. In the previous Sections, we stated that we perform an expansion in $\hbar$, in line with the familiar notion that the (semi)classical limit corresponds to $\hbar \to 0$. However, since $\hbar$ is itself a constant, this informal expression actually means that $\hbar$ should be small with respect to the relevant system parameters. More precisely, one should define a dimensionless semiclassical parameter for the expansion.

When converting Eq.~\eqref{eqn:operator-eigenvalue-diag} into an equation for symbols, one typically assumes that all terms in the Hamiltonian~$\hat{\cal H}$ are of the same order of magnitude. This energy scale can be denoted by $p_0^2/2m^*$,  where $p_0$ is the typical electron momentum, cf. Eq.~\eqref{eqn:valence_hamiltonian2}. 
Moreover, one assumes that the length scale $L$ of variations in the potentials, that is, in $\bm{\Delta}(\bm{r})$ and $\Delta_0(\bm{r})$ in Eq.~\eqref{eqn:valence_hamiltonian2}, is much larger than the length scale $\hbar/p_0$ set by the typical momentum $p_0$. Defining $\overline{\bm{r}} = \bm{r}/L$ and $\overline{\bm{\Delta}}(\overline{\bm{r}}) = 2m^* \bm{\Delta}(\bm{r})/p_0^2$, with a similar definition for $\Delta_0$, one then obtains the dimensionless Hamiltonian
\begin{equation}\label{eqn:valence_hamiltonian2-dimless}
  \overline{{\cal H}} \equiv \frac{2m^*{\cal H}}{p_0^2} = h^2 \frac{\partial^2}{\partial \overline{\bm{r}}^2} \tau_0 + \overline{\bm{\Delta}}(\overline{\bm{r}})\cdot\bm{\tau} + \overline{\Delta}_{0}(\overline{\bm{r}})\tau_0~,
\end{equation}
which contains the dimensionless semiclassical parameter $h=\hbar/(p_0 L)$.

Equation~\eqref{eqn:valence_hamiltonian2-dimless} makes it clear that our semiclassical expansion is actually an expansion in the parameter $h$. Unfortunately, this parameter is not small in our case. First, the typical electron momentum is on the order of $\hbar/\sqrt{A_{\rm M}}$, as it lies in the mini Brillouin zone. At the same time, $L$ can be taken as $\sqrt{A_{\rm M}}$, leading to $h \approx 1$.  This implies that we should regard a semiclassical expansion in $\hbar$ as a formal power series in the dimensionless semiclassical parameter, rather than an asymptotic power series.
One may remark that is related to a deeper problem, namely that not all terms in the Hamiltonian are of the same order. Indeed, as noted in Ref.~\cite{SM_morales-duran_PRL_2024}, the typical value $\Delta$ of $|\bm{\Delta}(\bm{r})|$ is much larger than $p_0^2/2m^*$.
This shows that the true small parameter in our problem is $p_0^2/(m^* \Delta) \approx \hbar^2/(m^* A_{\rm M} \Delta)\equiv\zeta$, which we call the relative energy splitting for reasons that will become clear shortly.
Unfortunately, defining a dimensionless semiclassical parameter $h$ using $\Delta$ instead of $p_0$ does not solve our problems with the expansion parameter. In some sense, it makes them worse, since it leads to problems with the definition of the momentum variable in the symbols. We therefore perform a semiclassical expansion in the parameter $\hbar$, or, more precisely, in $h = \hbar/p_0 L \approx 1$, regarding it as a formal expansion.
Consequently, we only gather all terms that are of the same order in the relative band splitting $\zeta$ at the end of our analysis.

We now proceed with solving  Eq.~(\ref{eqn:operator-eigenvalue-diag}) order by order in $\hbar$ for the Hamiltonian~(\ref{eqn:valence_hamiltonian2}).
As we discussed in the previous Section, we first use Eq.~(\ref{eqn:symbol-operator-product}) to express the symbols of the operator products in terms of the symbols of the individual operators. Taking Eq.~(\ref{eqn:symbol-operator-product-leading}), we have, to lowest order in $\hbar$~\cite{SM_Belov06,SM_Littlejohn91},
\begin{equation}  \label{eqn:symbol-eigenvalue-diag-pre}
  \mathcal{H}(\bm{r},\bm{p}) \chi^\pm(\bm{r},\bm{p},\hbar) = \chi^\pm(\bm{r},\bm{p},\hbar) H^\pm(\bm{r},\bm{p},\hbar) + \mathcal{O}(\hbar)~,
\end{equation}
where the symbol $\mathcal{H}(\bm{r},\bm{p})$ of $\hat{\cal H}$ is given by Eq.~(\ref{eqn:valence_hamiltonian2-symbol}), and where $\chi^\pm(\bm{r},\bm{p},\hbar)$ is the symbol of $\hat{\chi}$ and $H^\pm(\bm{r},\bm{p},\hbar)$ is the symbol of $\hat{H}^\pm$.
These last two symbols now have a power series expansion in $\hbar$, in the sense of Eq.~(\ref{eqn:symbol-expansion-powers}). Inserting these expansions into Eq.~(\ref{eqn:symbol-eigenvalue-diag-pre}), and retaining only the terms of $\mathcal{O}(\hbar^0)$, we obtain~\cite{SM_Belov06,SM_Littlejohn91}
\begin{equation}  \label{eqn:symbol-eigenvalue-diag}
  \mathcal{H}(\bm{r},\bm{p}) \chi_0^\pm(\bm{r},\bm{p}) = \chi_0^\pm(\bm{r},\bm{p}) H_0^\pm(\bm{r},\bm{p})~.
\end{equation}
In other words, the principal symbols $H_0^\pm$ and $\chi_0^\pm$ are given by the eigenvalues and eigenvectors, respectively, of $\mathcal{H}(\bm{r},\bm{p})$. 

Equation~\eqref{eqn:valence_hamiltonian2} shows that $\mathcal{H}(\bm{r},\bm{p})$ is diagonal in both the momentum operator and in $\Delta_0(\bm{r})$. The eigenvectors $\chi_0^\pm$ therefore correspond to the eigenvectors of $\bm{\Delta}(\bm{r})\cdot\bm{\tau}$, meaning that they are given by 
\begin{equation}  \label{eqn:eigenvectors-Delta-cdot-sigma}
  \chi_0^\pm(\bm{r},\bm{p}) = \frac{\sqrt{1 \mp n_z}}{\sqrt{2}} \begin{pmatrix} \pm \frac{n_x - i n_y}{1 \mp n_z} \\ 1 \end{pmatrix}~.
\end{equation}
In particular, the eigenvectors $\chi_0^\pm$ do not depend on the momentum coordinate $\bm{p}$. At the same time, the principal symbols $H_0^\pm$ are given by
\begin{equation}  \label{eqn:H-principal}
  H_0^\pm(\bm{r},\bm{p}) = -\frac{p^2}{2 m^*} + \Delta_0(\bm{r}) \pm | \bm{\Delta}(\bm{r}) |~.
\end{equation}
This shows that the energy splitting between the two pseudospin sectors is on the order of $2\Delta$, which is why we previously called $\zeta$ the relative energy splitting.
A couple of remarks are in order at this point. First of all, Eq.~(\ref{eqn:symbol-eigenvalue-diag}) does not require the eigenvectors $\chi_0^\pm$ to be normalized. This is, however, required by the conservation of probabilty (\ref{eqn:chi-unitary}). Converting the latter into an equation for the symbol $\chi(\bm{r},\bm{p},\hbar)$ and retaining only the lowest-order terms~\cite{SM_Belov06}, we directly obtain $(\chi_0^\pm)^\dagger \chi_0^\pm = 1$.
Second, we note that when combining the two eigenvectors $\chi_0^\pm$ into a matrix $U_0$, cf. Eq.~\eqref{eqn:operator-eigenvalue-diag-matrix}, we exactly obtain the matrix that is used in Refs.~\cite{SM_morales-duran_PRL_2024,SM_bruno_PRL_2004}. Thus, to lowest order, our semiclassical projection onto a single pseudospin sector is equivalent to the approach previously discussed in the literature. Our advantage is that we can also incorporate higher-order terms, as we will discuss shortly.
As a third remark, we note that our choice of the eigenvectors $\chi_0^\pm$ is not unique~\cite{SM_Littlejohn91,SM_Reijnders18}, as we may multiply them by $\exp(i g^\pm(\bm{r}))$, where $g^\pm(\bm{r})$ is an arbitrary (smooth) function. We will explore the consequences of this ``gauge freedom'' later on.

Using the product formula~(\ref{eqn:symbol-operator-product}) and the series expansion~(\ref{eqn:symbol-expansion-powers}) to convert Eq.~(\ref{eqn:operator-eigenvalue-diag}) into an equation for symbols, we can also construct~\cite{SM_Belov06,SM_Littlejohn91} all higher-order terms in the expansions of the symbols $H^\pm(\bm{r},\bm{p},\hbar)$ and $\chi^\pm(\bm{r},\bm{p},\hbar)$.
Collecting the terms of $\mathcal{O}(\hbar)$, we find
\begin{equation}  \label{eqn:symbol-eigenvalue-diag-O1}
  \mathcal{H} \chi_1^\pm + \frac{i}{2} \{ \mathcal{H}, \chi_0^\pm \} = \chi_0^\pm H_1^\pm + \chi_1^\pm H_0^\pm + \frac{i}{2} \{ \chi_0^\pm, H_0^\pm \}~,
\end{equation}
where we omitted the arguments $(\bm{r},\bm{p})$ of the various symbols, and used the fact that $\mathcal{H}$ only has a principal symbol. We now multiply this expression by $(\chi_0^\pm)^\dagger$, and use that $\chi_0^\pm$ is an eigenvector of $\mathcal{H}$, which implies that $(\chi_0^\pm)^\dagger (\mathcal{H} - H_0^\pm) = 0$. We then obtain~\cite{SM_Belov06,SM_Littlejohn91}
\begin{equation}
  H_1^\pm = \frac{i}{2} (\chi_0^\pm)^\dagger \{ \mathcal{H}, \chi_0^\pm \} - \frac{i}{2} (\chi_0^\pm)^\dagger \{ \chi_0^\pm, H_0^\pm \}~.
\end{equation}
Since $\mathcal{H}$ is diagonal in the momentum operator---see Eq.~(\ref{eqn:valence_hamiltonian2-symbol})---and $\chi_0^\pm$ does not depend on $\bm{p}$, we have
\begin{equation}  \label{eqn:L1-Poisson-bracket-simplification}
  \{ \mathcal{H}, \chi_0^\pm \} = - \left\langle \frac{\partial \mathcal{H}}{\partial \bm{p}} , \frac{\partial \chi_0^\pm}{\partial \bm{r}} \right\rangle 
  = - \{ \chi_0^\pm, H_0^\pm \}~.
\end{equation}
We therefore find that
\begin{equation}  \label{eqn:H1-general-Berry}
  H_1^\pm = - i (\chi_0^\pm)^\dagger \{ \chi_0^\pm, H_0^\pm \}~,
\end{equation}
which shows that $H_1^\pm$ has the character of a Berry connection~\cite{SM_Littlejohn91,SM_Reijnders18}. 
Using Eq.~(\ref{eqn:H-principal}) and the definition of the Poisson bracket, Eq.~(\ref{eqn:H1-general-Berry}) becomes
\begin{equation}  \label{eqn:H1-computed}
  \hbar H_1^\pm = i \hbar \sum_j \frac{p_j}{m^*} (\chi_0^\pm)^\dagger \frac{\partial \chi_0^\pm}{\partial r_j} = -\frac{1}{m^*} \left\langle \bm{p} , \frac{e}{c} \bm{A}^\pm \right\rangle~,
\end{equation}
where
\begin{equation}  \label{eqn:vector-potential}
  \bm{A}^{\pm} = \frac{\hbar c}{2e} \frac{1}{1 \mp n_z} \left( -n_x \frac{\partial n_y}{\partial \bm{r}} + n_y \frac{\partial n_x}{\partial \bm{r}}\right)~.
\end{equation}
In this last computation one uses the eigenvectors~\eqref{eqn:eigenvectors-Delta-cdot-sigma} and the fact that $\bm{n}(\bm{r})$ is a unit vector. Note that Eq.~\eqref{eqn:vector-potential} coincides with Eq.~\eqref{eqn:vector-potential-plus} in the main text for the upper layer-pseudospin sector.

We now briefly return to the freedom in the choice of $\chi_0^\pm$. As we previously mentioned, we can also consider a different eigenvector, namely
\begin{equation}  \label{eqn:chi0-gauge-transform}
  \tilde{\chi}_0^\pm(\bm{r}) = e^{ig^\pm(\bm{r})} \chi_0^\pm(\bm{r})~,
\end{equation}
where $g^\pm(\bm{r})$ is an arbitrary (smooth) function~\cite{SM_Littlejohn91,SM_Reijnders18}. Let us investigate how this affects the different symbols that we considered so far. First of all, the principal symbols $H_0^\pm$ do not change, since they are the eigenvalues in Eq.~(\ref{eqn:symbol-eigenvalue-diag}).
Starting from Eq.~(\ref{eqn:H1-general-Berry}), one can show~\cite{SM_Littlejohn91,SM_Reijnders18} that $H_1^\pm$ transforms as
$\tilde{H}_1^\pm = H_1^\pm + \{ g^\pm, H_0^\pm \}$.
With the help of Eq.~(\ref{eqn:H1-computed}), this becomes
\begin{equation}  \label{eqn:H1-gauge-final}
  \hbar \tilde{H}_1^\pm = -\frac{1}{m^*} \left\langle \bm{p} , \frac{e}{c} \bm{A}^\pm + \hbar \frac{\partial g^\pm}{\partial \bm{r}} \right\rangle~.
\end{equation}
We therefore see that the transformation~(\ref{eqn:chi0-gauge-transform}) adds a gradient term to the vector potential $\bm{A}^\pm$. Hence, we can indeed regard this transformation as a gauge transformation, since it does not affect the effective magnetic field $B_z^{\pm}$.

In many cases of physical interest~\cite{SM_Belov06,SM_Reijnders18}, it is sufficient to compute $H_0^\pm$, $H_1^\pm$ and $\chi_0^\pm$. However, in the present case we also have to consider $H_2^\pm$. This can be understood by noting that the vector potential $\bm{A}^\pm$ should enter the effective scalar Hamiltonian through minimal coupling. However, Eqs.~\eqref{eqn:H-principal} and~\eqref{eqn:H1-computed} are not sufficient to complete the square, since a term proportional to $|\bm{A}^\pm|^2$ is missing. Since the latter term is of order $\hbar^2$, see Eq.~\eqref{eqn:vector-potential}, it is part of $H_2^\pm$.
Before we can determine $H_2^\pm$, we first have to compute $\chi_1^\pm$. Since the eigenvectors $\chi_0^\pm$ form an orthonormal basis of the two-dimensional vectors, we may expand $\chi_1^\pm$ in this basis, i.e.
\begin{equation}  \label{eqn:chi1-basis-expansion}
  \chi_1^\alpha = c_{1,\alpha}^\alpha \chi_0^\alpha + c_{1,-\alpha}^\alpha \chi_0^{-\alpha}~,
\end{equation}
where $\alpha = \pm$ and $-\alpha = \mp$. Converting Eq.~(\ref{eqn:chi-unitary}) into an equation for symbols and collecting the terms of $\mathcal{O}(\hbar)$, we have
\begin{equation}
  (\chi_0^\alpha)^\dagger \chi_1^\alpha + (\chi_1^\alpha)^\dagger \chi_0^\alpha + \frac{i}{2} \{ (\chi_0^\alpha)^\dagger , \chi_0^\alpha \}~,
\end{equation}
where the last term vanishes because the eigenvectors $\chi_0^\alpha$ do not depend on $\bm{p}$. Inserting the basis expansion~(\ref{eqn:chi1-basis-expansion}), we have
$c_{1,\alpha}^\alpha + (c_{1,\alpha}^\alpha)^* = 0$, or $\mathrm{Re} \, c_{1,\alpha}^\alpha = 0$.
Our next step is to insert the basis expansion~(\ref{eqn:chi1-basis-expansion}) into Eq.~(\ref{eqn:symbol-eigenvalue-diag-O1}). We first observe that
\begin{equation}
  ( \mathcal{H} - H_0^\alpha ) \chi_1^\alpha = c^\alpha_{1,-\alpha} (H_0^{-\alpha} - H_0^\alpha) \chi_0^{-\alpha}~.
\end{equation}
When we also take Eq.~(\ref{eqn:L1-Poisson-bracket-simplification}) into account, Eq.~(\ref{eqn:symbol-eigenvalue-diag-O1}) becomes
\begin{equation}
  c^\alpha_{1,-\alpha} (H_0^{-\alpha} - H_0^\alpha) \chi_0^{-\alpha} = \chi_0^\alpha H_1^\alpha + i \{ \chi_0^\alpha, H_0^\alpha \}~.
\end{equation}
Multiplying by $(\chi_0^{-\alpha})^\dagger$, we then obtain
\begin{equation}  \label{eqn:c1-allpha-minus-alpha-general}
  c^\alpha_{1,-\alpha} = \frac{i}{H_0^{-\alpha} - H_0^\alpha} (\chi_0^{-\alpha})^\dagger \left\langle \frac{\partial \chi_0^\alpha}{\partial \bm{r}} , \frac{\partial H_0^\alpha}{\partial \bm{p}} \right\rangle~,
\end{equation}
where we also used that $\chi_0^\alpha$ does not depend on $\bm{p}$ to simplify the Poisson bracket. A brief computation then shows that
\begin{equation}  \label{eqn:c1-allpha-minus-alpha}
  c^\alpha_{1,-\alpha} = \frac{1}{4 |\bm{\Delta}|} \sum_j \frac{p_j}{m^*} \frac{1}{\sqrt{1-n_z^2}} \left( n_y \frac{\partial n_x}{\partial r_j} - n_x \frac{\partial n_y}{\partial r_j} - i \alpha \frac{\partial n_z}{\partial r_j} \right)~.
\end{equation}
Since our equations do not constrain the imaginary part of $c_{1,\alpha}^\alpha$, we may set it to zero, giving $c_{1,\alpha}^\alpha = 0$.
With this choice, to which we will come back momentarily, we have
\begin{equation}  \label{eqn:chi1-final}
  \chi_1^\alpha = c_{1,-\alpha}^\alpha \chi_0^{-\alpha}~.
\end{equation}
We observe that $\chi_1^\pm$ depends linearly on $\bm{p}$, contrary to $\chi_0^\pm$. Moreover, it is of first order in the relative energy splitting, that is, of order $\zeta$. This is in contrast to $\chi_0^\pm$, which is of zeroth order in this splitting and which was used in Refs.~\cite{SM_morales-duran_PRL_2024,SM_bruno_PRL_2004} to perform the projection onto a single pseudospin sector. We therefore conclude that $\chi_1^\pm$ represents the first higher-order term in our asymptotic expansion. We may say that $\chi_1^\pm$ can be used to diagonalize the off-diagonal terms that were neglected in Refs.~\cite{SM_morales-duran_PRL_2024,SM_bruno_PRL_2004} since they are of first order in the relative energy splitting. As we will see shortly, $\chi_1^\pm$ indeed gives rise to a term in $H_2$ that is of first order in the relative energy splitting.

Now that we have computed $H_0^\pm$, $H_1^\pm$, $\chi_0^\pm$ and $\chi_1^\pm$, we proceed with $H_2^\pm$.
To this end, we need the term of order $\hbar^2$ in the product formula Eq.~(\ref{eqn:symbol-operator-product}), which is given by
\begin{equation}
  -\frac{\hbar^2}{8} \sum_{j,k} \bigg( 
  \frac{\partial^2 a}{\partial p_j \partial p_k} \frac{\partial^2 b}{\partial r_j \partial r_k}
  - \frac{\partial^2 a}{\partial p_j \partial r_k} \frac{\partial^2 b}{\partial r_j \partial p_k}
  - \frac{\partial^2 a}{\partial r_j \partial p_k} \frac{\partial^2 b}{\partial p_j \partial r_k}
  + \frac{\partial^2 a}{\partial r_j \partial r_k} \frac{\partial^2 b}{\partial p_j \partial p_k}
  \bigg)~,
  \label{eqn:symbol-operator-product-second-order}
\end{equation}
where the second and third term are equal due to the equality of mixed partials.
Gathering all terms of $\mathcal{O}(\hbar^2)$ after converting Eq.~(\ref{eqn:operator-eigenvalue-diag}) into an equation for symbols, we obtain a large number of terms. Noting that most of the second derivative terms~(\ref{eqn:symbol-operator-product-second-order}) vanish because $\chi_0$ does not depend on $\bf{p}$, we find the equation
\begin{equation}
  \mathcal{H} \chi_2 + \frac{i}{2} \{ \mathcal{H} , \chi_1 \} 
  - \frac{1}{8} \sum_{j,k} \frac{\partial^2 \mathcal{H}}{\partial p_j \partial p_k} \frac{\partial^2 \chi_0}{\partial r_j \partial r_k} 
  = \chi_0 H_2 + \chi_1 H_1 + \chi_2 H_0 + \frac{i}{2} \{ \chi_1 , H_0 \} 
  + \frac{i}{2} \{ \chi_0 , H_1 \} 
  - \frac{1}{8} \sum_{j,k} \frac{\partial^2 H_0}{\partial p_j \partial p_k} \frac{\partial^2 \chi_0}{\partial r_j \partial r_k}~,
  \label{eqn:symbol-eigenvalue-diag-O2}
\end{equation}
where we omitted the superscripts $\pm$ to lighten the notation. Since the Hamiltonian $\mathcal{H}$ is diagonal in $\bm{p}$, the second derivatives of $\mathcal{H}$ and $H_0$ with respect to $\bm{p}$ coincide (and are in fact proportional to $\delta_{jk}$). Therefore, the terms with the second derivatives in Eq.~(\ref{eqn:symbol-eigenvalue-diag-O2}) cancel. We multiply the remaining terms by $\chi_0^\dagger$ from the left. Using that $\chi_0$ is an eigenvalue of $\mathcal{H}$ and using that $(\chi_0^\alpha)^\dagger \chi_1^\alpha = 0$ when $\chi_1^\alpha$ is given by Eq.~(\ref{eqn:chi1-final}), we have
\begin{equation}  \label{eqn:H2-semi-general}
  H_2 = -\frac{i}{2} \chi_0^\dagger \{ \chi_0 , H_1 \} + \frac{i}{2} \chi_0^\dagger \{ \mathcal{H} , \chi_1 \} - \frac{i}{2} \chi_0^\dagger \{ \chi_1 , H_0 \}~.
\end{equation}

We start by focusing on the first term in this expression. As $H_1$ is itself given by a Poisson bracket, see Eq.~\eqref{eqn:H1-general-Berry}, and proportional to the vector potential, see Eq.~\eqref{eqn:H1-computed}, we may intuitively expect that this term gives rise to the vector potential squared.
Indeed, a short computation shows that
\begin{equation}  \label{eqn:H2-semi-general-first-term}
  -\frac{i \hbar^2}{2} \chi_0^\dagger \{ \chi_0 , H_1 \} 
  = \frac{i \hbar}{2 m^*} \left\langle \chi_0^\dagger \frac{\partial \chi_0}{\partial \bm{r}} , \frac{e}{c} \bm{A} \right\rangle
  = - \frac{e^2 |\bm{A}|^2}{2 m^* c^2}~,
\end{equation}
cf. Eq.~\eqref{eqn:H1-computed}. We briefly verify that the gauge transformation~\eqref{eqn:chi0-gauge-transform} transforms this term in the correct way. A brief calculation using our previous result~(\ref{eqn:H1-gauge-final}) shows that
\begin{equation}
  -\frac{i \hbar^2}{2} \tilde{\chi}_0^\dagger \{ \tilde{\chi}_0 , \tilde{H_1} \} = -\frac{1}{2 m^*} \left| \frac{e}{c} \bm{A} + \hbar \frac{\partial g_0}{\partial \bm{r}} \right|^2~.
\end{equation}
Hence, the transformation~\eqref{eqn:chi0-gauge-transform} adds the same gradient term to the vector potential $\bm{A}$ as in the linear term~\eqref{eqn:H1-computed}. This allows us to complete the square later on, and shows that we are indeed dealing with a gauge transformation.

We proceed with the second term in Eq.~(\ref{eqn:H2-semi-general}). Using the definition of the Poisson bracket, we have
\begin{equation}
  \chi_0^\dagger \{ \mathcal{H} , \chi_1 \}
  = \chi_0^\dagger \left\langle \frac{\partial \mathcal{H}}{\partial \bm{r}} , \frac{\partial \chi_1}{\partial \bm{p}} \right\rangle - \chi_0^\dagger \left\langle \frac{\partial \mathcal{H}}{\partial \bm{p}} , \frac{\partial \chi_1}{\partial \bm{r}} \right\rangle~.
\end{equation}
Taking the derivative of Eq.~(\ref{eqn:symbol-eigenvalue-diag}) with respect to $\bm{r}$ and $\bm{p}$, we can eliminate the derivatives of $\mathcal{H}$ from the above expression. Noting that $\chi_0$ does not depend on $\bm{p}$, we obtain
\begin{equation}  \label{eqn:H2-semi-general-second-term}
  \chi_0^\dagger \{ \mathcal{H} , \chi_1 \}
  = \bigg\langle \frac{\partial \chi_0^\dagger}{\partial \bm{r}} (H_0 - \mathcal{H}) , \frac{\partial \chi_1}{\partial \bm{p}} \bigg\rangle 
  - \chi_0^\dagger \{ \chi_1 , H_0 \}~.
\end{equation}
The second term in this result can be combined with the third term in Eq.~(\ref{eqn:H2-semi-general}). Using Eq.~(\ref{eqn:chi1-final}), the first term in Eq.~\eqref{eqn:H2-semi-general-second-term} becomes
\begin{equation}  \label{eqn:H2-semi-general-second-term-first-part}
  \bigg\langle \frac{\partial (\chi_0^\alpha)^\dagger}{\partial \bm{r}} (H_0^\alpha - \mathcal{H}) , \frac{\partial \chi_1^\alpha}{\partial \bm{p}} \bigg\rangle 
  = \sum_j \frac{\partial (\chi_0^\alpha)^\dagger}{\partial r_j} (H_0^\alpha - H_0^{-\alpha}) \chi_0^{-\alpha} \frac{\partial c_{1,-\alpha}^\alpha}{\partial p_j}~.
\end{equation}
We note that $H_0^\alpha - H_0^{-\alpha} = 2\alpha|\bm{\Delta}|$, which cancels the factor $|\bm{\Delta}|^{-1}$ in $c_{1,-\alpha}^\alpha$, see Eq.~\eqref{eqn:c1-allpha-minus-alpha}. The term~\eqref{eqn:H2-semi-general-second-term-first-part} therefore gives rise to a contribution of leading order in the relative energy splitting $\zeta$, even though $\chi_1$ is of first order in this parameter.
Inserting the general expression~(\ref{eqn:c1-allpha-minus-alpha-general}) for $c_{1,-\alpha}^\alpha$, and performing some algebraic manipulations analogous to the ones which lead to Eq.~\eqref{eqn:c1-allpha-minus-alpha}, we obtain
\begin{equation}
  \bigg\langle \frac{\partial (\chi_0^\alpha)^\dagger}{\partial \bm{r}} (H_0^\alpha - \mathcal{H}) , \frac{\partial \chi_1^\alpha}{\partial \bm{p}} \bigg\rangle 
  = \frac{i}{m^*} \sum_j\left| (\chi_0^{-\alpha})^\dagger \frac{\partial \chi_0^\alpha}{\partial r_j} \right|^2
  = \frac{i}{4 m^*} \sum_j (\partial_{r_j} \bm{n}) \cdot (\partial_{r_j} \bm{n})~,
  \label{eqn:H2-D}
\end{equation}
where the dot indicates the inner product over the components of the vector $\bm{n}$.

We now compute the third and last term in Eq.~(\ref{eqn:H2-semi-general}), which we combine with the second term in Eq.~(\ref{eqn:H2-semi-general-second-term}). 
Using the properties of Poisson brackets, we have
\begin{equation}
  (\chi_0^\alpha)^\dagger \{ \chi_1^\alpha , H_0^\alpha \}
  = \{ (\chi_0^\alpha)^\dagger \chi_1^\alpha , H_0^\alpha \} - \{ (\chi_0^\alpha)^\dagger , H_0^\alpha \} \chi_1^\alpha 
  = - \left\langle \frac{\partial (\chi_0^\alpha)^\dagger}{\partial \bm{r}} , \frac{\partial H_0^\alpha}{\partial \bm{p}} \right\rangle \chi_1^\alpha~,
\end{equation}
where the last equality follows from Eq.~(\ref{eqn:chi1-final}), and the fact that $\chi_0^\alpha$ does not depend on $\bm{p}$. Inserting Eqs.~(\ref{eqn:H-principal}),~(\ref{eqn:c1-allpha-minus-alpha-general}) and~(\ref{eqn:chi1-final}), and performing some algebraic manipulations, we have
\begin{equation}
  (\chi_0^\alpha)^\dagger \{ \chi_1^\alpha , H_0^\alpha \} 
  = - \frac{i}{H_0^{-\alpha} - H_0^\alpha} \sum_{j,k} \frac{p_j p_k}{(m^*)^2}
  \bigg((\chi_0^{-\alpha})^\dagger \frac{\partial \chi_0^\alpha}{\partial r_j}\bigg)^\dagger
  \bigg((\chi_0^{-\alpha})^\dagger \frac{\partial \chi_0^\alpha}{\partial r_k}\bigg)~.
\end{equation}
The terms with $j=k$ in this sum are similar to the terms in Eq.~(\ref{eqn:H2-D}), and follow straightforwardly.
The off-diagonal terms can be computed in a similar fashion, yielding the final result
\begin{equation}  \label{eqn:H2-higher-order}
  (\chi_0^\alpha)^\dagger \{ \chi_1^\alpha , H_0^\alpha \} 
  = - \frac{i}{H_0^{-\alpha} - H_0^\alpha} \sum_{j,k} \frac{p_j p_k}{(m^*)^2} \frac{1}{4} (\partial_{r_j} \bm{n}) \cdot (\partial_{r_k} \bm{n}) 
  = \frac{i \alpha}{8 (m^*)^2 | \bm{\Delta} |} \left| \left\langle \bm{p}, \frac{\partial \bm{n}}{\partial \bm{r}} \right\rangle \right|^2~,
\end{equation}
where $|..|^2$ indicates the inner product over the components of the vector $\bm{n}$. When we multiply this term by $\hbar^2$, we find that the result is of order $\zeta p_0^2/m^*$. It is therefore of first order in the relative energy splitting.

Combining Eqs.~\eqref{eqn:H2-semi-general}, \eqref{eqn:H2-semi-general-first-term}, \eqref{eqn:H2-D} and~\eqref{eqn:H2-higher-order}, we finally obtain
\begin{equation}  \label{eqn:H2-computed}
  \hbar^2 H_2^\alpha = - \frac{e^2 |\bm{A}^\alpha(\bm{r})|^2}{2 m^* c^2} - D(\bm{r}) + \frac{\alpha\hbar^2}{8 (m^*)^2 | \bm{\Delta}(\bm{r}) |} \left| \left\langle \bm{p}, \frac{\partial \bm{n}}{\partial \bm{r}} \right\rangle \right|^2~,
\end{equation}
where $D(\bm{r})$ was defined below Eq.~\eqref{eqn:magnetic-field-z} in the main text, and $\alpha=\pm$ as before. In the computations that lead to this result, we set $c_{1,\alpha}^\alpha$ to zero. However, since its imaginary parts is not constrained by our equations, we could have also made a different choice. We may therefore also consider
\begin{equation}  \label{eqn:chi1-gauge-transform}
  \tilde{\chi}_1^\alpha 
  = \chi_1^\alpha + i g_1^\alpha(\bm{r}) \chi_0^\alpha
  = c_{1,-\alpha}^\alpha \chi_0^{-\alpha} + i g_{1}^{\alpha}(\bm{r}) \chi_0^\alpha~,
\end{equation}
where $g_{1}^{\alpha}(\bm{r})$ is an arbitrary (smooth) function. Repeating the various steps in the derivation, we find that this leads to
\begin{equation}
  \tilde{H}_2 
  = H_2 - i g_1 H_1 + g_1 \chi_0^\dagger \{ \chi_0 , H_0 \} + \{ g_1 , H_0 \}
  = H_2 - \frac{1}{m^*} \left\langle \bm{p} , \frac{\partial g_1}{\partial \bm{r}} \right\rangle~.
\end{equation}
In other words, the transformation~\eqref{eqn:chi1-gauge-transform} adds a gradient term to the vector potential, which does not affect the magnetic field. We can therefore regard it as another gauge transformation that does not alter our result.

Adding the results~(\ref{eqn:H-principal}),~(\ref{eqn:H1-computed}) and~(\ref{eqn:H2-computed}), and completing the square, we find the symbol of the effective scalar Hamiltonian corresponding to a single layer-pseudospin sector as
\begin{equation}
  H^{\pm} = -\frac{1}{2m^*}\left(\bm{p} + \frac{e}{c}\bm{A}^{\pm}\right)^2 + \Delta_0(\bm{r}) \pm | \bm{\Delta}(\bm{r}) | 
  - D(\bm{r})
  \pm \frac{\hbar^2}{8 (m^*)^2 | \bm{\Delta} |} \left| \left\langle \bm{p}, \frac{\partial \bm{n}}{\partial \bm{r}} \right\rangle \right|^2 + \mathcal{O}(\hbar^2)~.
  \label{eqn:adiabatic-Hamiltonian-single-band}
\end{equation}
We now return to our discussion of the different expansion parameters, which we anticipated at the beginning of this Section. So far, we have performed a formal series expansion up to second order in $\hbar$, or, more precisely, in $\hbar/(p_0 L)$. As we showed, this parameter is not actually small, which explains why we can combine the zeroth-order term $\bm{p}^2$ with the first-order term $\langle \bm{p}, \bm{A} \rangle$ in the final result~\eqref{eqn:adiabatic-Hamiltonian-single-band}. 
The relative energy splitting $\zeta$ is, however, a small parameter. We therefore conclude that the last term in Eq.~\eqref{eqn:adiabatic-Hamiltonian-single-band} is a higher-order correction that can be neglected when we are only interested in the leading-order term of the effective scalar Hamiltonian. Comparing Eq.~\eqref{eqn:adiabatic-Hamiltonian-single-band} to the result in Refs.~\cite{SM_morales-duran_PRL_2024,SM_bruno_PRL_2004}, we observe that we have obtained all contributions to this leading-order term, which means that we do not have to compute $H_3$, $H_4$, and so forth. At the same time, we emphasize that computing these terms provides a systematic way to construct all terms in the effective Hamiltonian that are of higher order in the relative energy splitting $\zeta$. The latter terms were neglected in the previous derivations~\cite{SM_morales-duran_PRL_2024,SM_bruno_PRL_2004} at the point where the off-diagonal terms are neglected in favor of the diagonal terms after conjugating the Hamiltonian with $U_0$.

The final step in our approach is to obtain the effective quantum Hamiltonian $\hat{H}^\pm$ corresponding to a single layer-pseudospin sector from its symbol~\eqref{eqn:adiabatic-Hamiltonian-single-band}. The quantization procedure~\eqref{eqn:quantization-Weyl} is particularly simple in this case, since most terms only depend on either $\bm{r}$ or $\bm{p}$. After neglecting the last term, which is proportional to $\zeta$, the only term that contains a product of $\bm{r}$ and $\bm{p}$ is $H_1^\pm$, given by Eq.~\eqref{eqn:H1-computed}. Since this term has the form discussed below Eq.~(\ref{eqn:valence_hamiltonian2-symbol}), we can straightforwardly quantize it.
In this way, we obtain
\begin{equation}
  \hat{H}^{\pm} = -\frac{1}{2m^*}\left(\hat{\bm{p}} + \frac{e}{c}\bm{A}^{\pm}\right)^2 + \Delta_0(\bm{r}) \pm | \bm{\Delta}(\bm{r}) | 
  - D(\bm{r})~,
  \label{eqn:adiabatic-Hamiltonian-single-band-quantum}
\end{equation}
which coincides with previous results~\cite{SM_morales-duran_PRL_2024,SM_bruno_PRL_2004} and corresponds to Eq.~\eqref{eqn:Hamiltonian-effective-single-band} in the main text when considering the upper layer-pseudospin sector.

\section*{Section IV: Conductivity and polarization function in the Landau-level model}

In this Section, we analytically compute the inter-band optical conductivity for the transition between the first and second valence bands of the Landau-level Hamiltonian~\eqref{eqn:H-Landau-levels}, and show that it leads to the plasmon dispersion~\eqref{eqn:magnetoplasmon}. We also calculate the full polarization function for the same transition, and show that it leads to the same result for the plasmon dispersion in the limit of small momenta.

For a generic system with discrete energy levels $\epsilon_\lambda$ and eigenstates $|\lambda\rangle$, we can write down the inter-band conductivity as~\cite{SM_GiulianiVignale}
\begin{equation}  \label{eqn:conductivity-interband-general}
  \sigma_{\alpha\beta}^{\rm inter}(\omega) = -\frac{i e^2 \hbar}{A} \sum_{\lambda' \neq \lambda} \frac{f_{\lambda'} - f_\lambda}{\epsilon_{\lambda'} - \epsilon_\lambda} \frac{{\cal M}_{\lambda'\lambda,\alpha\beta}}{\epsilon_{\lambda'}-\epsilon_\lambda + \hbar\omega +i\eta}~,
\end{equation}
where $A$ is the area of the system and $f_{\lambda}$ is the occupation of state $\lambda$. The matrix element ${\cal M}_{\lambda'\lambda,\alpha\beta}$ is defined by 
\begin{equation}
  {\cal M}_{\lambda'\lambda,\alpha\beta} = \langle \lambda' | \hat{v}_\alpha | \lambda \rangle \langle \lambda | \hat{v}_\beta | \lambda' \rangle~,
\end{equation}
where $\hat{v}_\alpha$ represents the velocity operator.
Splitting the sum in Eq.~\eqref{eqn:conductivity-interband-general} into two separate sums involving $f_{\lambda'}$ and $f_\lambda$, and interchanging the indices $\lambda'$ and $\lambda$ in the second sum, we obtain, cf. Eq.~\eqref{eqn:SM_conductivity_inter-band}
\begin{align}
  \sigma_{\alpha\beta}^{\rm inter}(\omega) 
    &= -\frac{i e^2 \hbar}{A} \sum_{\lambda' \neq \lambda} \frac{1}{\epsilon_{\lambda'} - \epsilon_\lambda} \left( 
    f_{\lambda'} \frac{{\cal M}_{\lambda'\lambda,\alpha\beta}}{\epsilon_{\lambda'}-\epsilon_\lambda + \hbar\omega +i\eta} 
    + f_{\lambda'} \frac{{\cal M}_{\lambda\lambda',\alpha\beta}}{\epsilon_\lambda-\epsilon_{\lambda'} + \hbar\omega +i\eta} \right)~, \nonumber \\
    &= -\frac{i e^2 \hbar}{A} \sum_{\lambda' \neq \lambda} \frac{f_{\lambda'}}{\epsilon_{\lambda'} - \epsilon_\lambda} \left( 
    \frac{{\cal M}_{\lambda'\lambda,\alpha\beta}}{\epsilon_{\lambda'}-\epsilon_\lambda + \hbar\omega +i\eta} 
    - \frac{{\cal M}_{\lambda'\lambda,\alpha\beta}^*}{\epsilon_{\lambda'}-\epsilon_\lambda - \hbar\omega - i\eta} \right)~, \nonumber \\
    &= -\frac{i e^2 \hbar}{A} \sum_{\lambda' \neq \lambda} \frac{f_{\lambda'}}{\epsilon_{\lambda'} - \epsilon_\lambda} 
    \frac{2 i (\epsilon_{\lambda'}-\epsilon_\lambda) {\rm Im} \, {\cal M}_{\lambda'\lambda,\alpha\beta} - 2 \hbar\omega {\rm Re} \, {\cal M}_{\lambda'\lambda,\alpha\beta}}{(\epsilon_{\lambda'}-\epsilon_\lambda)^2 - (\hbar\omega + i\eta)^2}~,
    \label{eqn:condictivity-general-n-levels}
\end{align}
where we used that ${\cal M}_{\lambda\lambda',\alpha\beta} = {\cal M}_{\lambda'\lambda,\alpha\beta}^*$ in the second equality.

For the Landau-level Hamiltonian~\eqref{eqn:H-Landau-levels}, with $\hat{H}_{\rm cor} = 0$, the eigenstates in the Landau gauge are given by $|\lambda\rangle = |n,k_y\rangle e^{i k_y y}/\sqrt{L_y}$, with $L_y$ the length of the sample in the $y$-direction. Using the relations
\begin{equation}
  \hat{v}_x = -\frac{1}{m^*} \hat{\Pi}_x = -\frac{1}{m^*} \frac{i\hbar}{\sqrt{2}\ell} (\hat{a}^\dagger - \hat{a})~,
  \qquad
  \hat{v}_y = -\frac{1}{m^*} \hat{\Pi}_y = -\frac{1}{m^*} \frac{\hbar}{\sqrt{2}\ell} (\hat{a}^\dagger + \hat{a})~,
\end{equation}
which can be derived from Eq.~\eqref{eqn:def-ladder-operators}, and the properties of the ladder operators, i.e.,
\begin{equation}
  \hat{a}^\dagger |n,k_y\rangle = \sqrt{n+1} |n+1,k_y\rangle , \qquad \hat{a} |n,k_y\rangle = \sqrt{n} |n-1,k_y\rangle~,
\end{equation}
one straightforwardly computes the matrix elements
\begin{align}
  \langle m | \hat{v}_x | n \rangle
    &= \frac{1}{L_y} \int d y \, e^{i(k_y-k_y')y} \left\langle m,k_y' \left| -\frac{i\hbar}{\sqrt{2}m^*\ell} (\hat{a}^\dagger - \hat{a}) \right| n, k_y \right\rangle
    = - \frac{i\hbar}{\sqrt{2}m^*\ell} \big( \sqrt{n+1}\delta_{m,n+1} - \sqrt{n} \delta_{m,n-1} \big) \delta_{k_y,k_y'}~,
  \nonumber \\
  \langle m | \hat{v}_y | n \rangle 
    &= \frac{1}{L_y} \int d y \, e^{i(k_y-k_y')y} \left\langle m,k_y' \left| -\frac{\hbar}{\sqrt{2}m^*\ell} (\hat{a}^\dagger + \hat{a}) \right| n, k_y \right\rangle
    = -\frac{\hbar}{\sqrt{2}m^*\ell} \big( \sqrt{n+1}\delta_{m,n+1} + \sqrt{n} \delta_{m,n-1} \big) \delta_{k_y,k_y'}~.
    \label{eqn:matrix-elements-Landau}
\end{align}
Noting that $\epsilon_n=-(n+\tfrac{1}{2})\hbar\omega_{\rm c}$, one finds that the conductivity is given by
\begin{equation}  \label{eqn:conductivity-Landau-level-intermediate}
  \sigma_{\alpha\beta}^{\rm inter}(\omega) 
  = -\frac{i e^2 \hbar}{A} \sum_{m \neq n,k_y,k_y'} \frac{f_m}{(n-m)\hbar\omega_{\rm c}} 
  \frac{2 i (n-m)\hbar\omega_{\rm c} {\rm Im} \, {\cal M}_{mn,\alpha\beta} - 2 \hbar\omega {\rm Re} \, {\cal M}_{mn,\alpha\beta}}{(n-m)^2\hbar^2\omega_{\rm c}^2 - (\hbar\omega + i\eta)^2}~.
\end{equation}
As noted in Ref.~\cite{SM_GiulianiVignale}, the matrix elements~\eqref{eqn:matrix-elements-Landau} only depend on $k_y$ and $k_y'$ through a delta function. The sum over these variables in Eq.~\eqref{eqn:conductivity-Landau-level-intermediate} therefore reduces to a multiplication by $A/(2\pi\ell^2)$.

At this point, we note that our system has only two levels. The first valence band ($m=0$) is empty, so $f_0$ vanishes, and the second valence band ($m=1$) is completely filled, so $f_1=1$.
Performing the summation over $m\neq n \in \{0,1\}$, we find that
\begin{equation} 
  \sigma_{\alpha\beta}^{\rm inter}(\omega) =
    \frac{i e^2 \hbar}{2\pi\ell^2} \frac{1}{\hbar\omega_{\rm c}}
    \frac{\hbar^2}{2(m^*)^2\ell^2}
    \frac{1}{\hbar^2\omega_{\rm c}^2 - (\hbar^2\omega + i\eta)^2}
    \begin{pmatrix}
      -2\hbar\omega & -2i\hbar\omega_{\rm c} \\
      2i\hbar\omega_{\rm c} & -2\hbar\omega
    \end{pmatrix}~.
\end{equation}
We can simplify this expression by noting that $A_{\rm M} = 2\pi\ell^2$ and $\hbar\omega_{\rm c} = \hbar^2/(m^*\ell^2)$, which follows from the definitions in the main text. This yields
\begin{equation}
  \sigma_{\alpha\beta}^{\rm inter}(\omega) =
  \frac{e^2}{m^* A_{\rm M}}
  \frac{1}{\omega_{\rm c}^2 - (\omega + i\eta)^2}
  \begin{pmatrix}
    -i\omega & \omega_{\rm c} \\
    -\omega_{\rm c} & -i\omega
  \end{pmatrix}~.
\end{equation}
Since the total number of electrons in a single Landau level equals $N=A/(2\pi\ell^2)=A/A_{\rm M}$, see e.g. Ref.~\cite{SM_GiulianiVignale}, the electron density is given by $n_0=N/A=1/A_{\rm M}$. We therefore obtain
\begin{equation}  \label{eqn:conductivity-Landau-level-model-SM}
  \sigma_{\alpha\beta}^{\rm inter}(\omega) =
  \frac{n_0 e^2}{m^*} \frac{1}{\omega_{\rm c}^2 - (\omega + i\eta)^2} 
  \begin{pmatrix}
    -i\omega & \omega_{\rm c} \\
    -\omega_{\rm c} & -i\omega
  \end{pmatrix}~,
\end{equation}
which corresponds to Eq.~\eqref{eqn:conductivity-Landau-level-model} in the main text.
We obtain Eq.~\eqref{eqn:magnetoplasmon} for the plasmon dispersion in the long-wavelength limit ${\bm q} \to {\bm 0}$ by inserting the result~\eqref{eqn:conductivity-Landau-level-model-SM} into Eq.~\eqref{eqn:dielectric_tensor} for the dynamical dielectric function and by subsequently demanding that its real part vanishes.

We remark that we can also cast the plasmon dispersion in a form that is similar to Eqs.~\eqref{eqn:plasmon_dispersion} and~\eqref{eqn:inter-band_velocity}.
Comparing Eq.~\eqref{eqn:condictivity-general-n-levels} with Eq.~\eqref{eqn:SM_conductivity_inter-band} and noting that ${\cal M}_{mn,xx}$ is real, we can express the longitudinal inter-band conductivity as
\begin{equation}  \label{eq:conductivity-Landau-level-xx-alpha}
  \sigma_{xx}^{\rm inter}(\omega) = -\frac{2 i e^2}{\hbar} \hbar^2\alpha_{\rm L} \frac{\omega}{\omega_{\rm c}} \frac{1}{\hbar^2\omega_{\rm c}^2 - (\hbar\omega + i\eta)^2}, \qquad 
  \hbar^2\alpha_{\rm L} = -\frac{\hbar^2}{A} \sum_{m\neq n \in \{0,1\}} \sum_{k_y,k_y'} (-1)^m f_m {\cal M}_{mn,xx}~.
\end{equation}
These expressions are very similar to Eqs.~\eqref{eqn:alpha} and~\eqref{eqn:conductivity_SM}, with the main difference that $\omega_{01}$ has been replaced by $\omega_{\rm c}$ and that the definition of $\alpha$ has been adapted to the system at hand.
Inserting Eq.~\eqref{eq:conductivity-Landau-level-xx-alpha} into Eq.~\eqref{eqn:dielectric_tensor}, we find the plasmon dispersion
\begin{equation}  \label{eqn:plasmon_dispersion-Landau-level-model}
  \omega_{\rm pl}(q) = \sqrt{\omega_{\rm c}^2 + \frac{4\pi e^2}{\bar{\varepsilon}\hbar\omega_{\rm c}} \alpha_{\rm L} q}~,
\end{equation}
which corresponds to Eq.~\eqref{eqn:plasmon_dispersion} with $\omega_{01}$ replaced by $\omega_{\rm c}$ and $\alpha_{K}+\alpha_{K^\prime}$ by $\alpha_{\rm L}$.
A brief calculation using the matrix elements~\eqref{eqn:matrix-elements-Landau} shows that $\alpha_{\rm L} = \hbar \omega_{\rm c}/(4\pi m^* \ell^2)$. With the relation $n=1/A_{\rm M}=1/(2\pi\ell^2)$, one subsequently recovers the plasmon dispersion~\eqref{eqn:magnetoplasmon} from Eq.~\eqref{eqn:plasmon_dispersion-Landau-level-model}.

Until this point, we computed the dynamical dielectric function $\varepsilon({\bm q},\omega)$ from the inter-band conductivity through Eq.~\eqref{eqn:dielectric_tensor}, which yields a result that is valid in the long-wavelength limit ${\bm q} \to {\bm 0}$. However, for the Landau-level model, given by Eq.~\eqref{eqn:H-Landau-levels} with $\hat{H}_{\rm cor}=0$, we can also compute the dielectric function by computing the polarization $\chi({\bm q},\omega)$. In this way, we obtain a result that is valid for all momenta ${\bm q}$.

Taking only two Landau levels into account, and including spin degeneracy, we have~\cite{SM_GiulianiVignale}
\begin{equation}
  \chi_0({\bm q},\omega) = \frac{1}{A} \sum_{m\neq n\in \{0,1\}} \sum_{k_y,k_y'} \frac{f_m - f_{n}}{\epsilon_{m} - \epsilon_{n} + \hbar\omega + i\eta} \left| \left\langle n,k_y' | \hat{n}_{\bm q} | m,k_y \right\rangle \right|^2~,
  \label{eqn:chi0-LL-GV}
\end{equation}
where $\hat{n}_{\bm q}$ is the density operator. The matrix element can be computed explicitly~\cite{SM_GiulianiVignale}:
\begin{align}\label{eqn:matrix-el-Landau-def-F}
  \left\langle n,k_y' | \hat{n}_{\bm q} | m,k_y \right\rangle = e^{-i q_x(k_y+k_y')\ell^2/2 } F_{n m}(\bm{q}) \delta_{k_y-k_y',q_y}~,
\end{align}
where 
\begin{equation}
F_{n m}(\bm{q}) = \sqrt{\frac{m!}{n!}} \left( \frac{(q_x-iq_y)\ell}{\sqrt{2}} \right)^{n-m} e^{-q^2\ell^2/4} L_m^{n-m}\left(\frac{q^2\ell^2}{2}\right)
\end{equation}
for $n \geq m$ and $F_{n m}(\bm{q}) = (F_{m n}(-\bm{q}))^*$ for $n<m$.
The functions $L_{m}^n(x)$ are the associated Laguerre polynomials.
The summation over $k_y'$ in Eq.~\eqref{eqn:chi0-LL-GV} is straightforward because of the delta function. Since the resulting expression does not depend on $k_y$, the summation over this variable becomes~\cite{SM_GiulianiVignale} a multiplication by $A/(2\pi\ell^2)$.
In the end, we have
\begin{equation}  \label{eqn:pol-full-pre1}
  \chi_0({\bm q},\omega) = \frac{1}{2\pi\ell^2} \sum_{m\neq n\in \{0,1\}} \frac{f_m - f_{n}}{\epsilon_{m} - \epsilon_{n} + \hbar\omega + i\eta} \left| F_{n m}(\bm{q}) \right|^2~.
\end{equation}
Noting that $|F_{n m}|^2$ only depends on $|{\bm q}|$ and taking its definition into account, we conclude that $|F_{n m}(\bm{q})|^2 = |F_{m n}(\bm{q})|^2$. Splitting the sum into two parts and interchanging the indices in the same way as before, we can rewrite Eq.~\eqref{eqn:pol-full-pre1} as
\begin{equation}
  \chi_0({\bm q},\omega) = \frac{2}{2\pi\ell^2} \sum_{m\neq n\in \{0,1\}} f_m \frac{\epsilon_m - \epsilon_{n}}{(\epsilon_m - \epsilon_{n})^2 - (\hbar\omega + i\eta)^2} \left| F_{n m}(\bm{q}) \right|^2~.
\end{equation}
This sum can easily be computed, since we have only two levels, of which only one is occupied. Taking into account that $|F_{01}(\bm{q})|^2 = (q^2\ell^2/2) \exp(-q^2 \ell^2/2)$ and using that  $\epsilon_n=-(n+\tfrac{1}{2})\hbar\omega_{\rm c}$, we arrive at
\begin{equation}
  \chi_0({\bm q},\omega) = -\frac{q^2}{2\pi\hbar} \frac{\omega_{\rm c}}{\omega_{\rm c}^2 - (\omega + i\eta)^2} \exp(-q^2 \ell^2/2)
    =-\frac{q^2}{m^* A_M} \frac{1}{\omega_{\rm c}^2 - (\omega + i\eta)^2} \exp(-q^2 \ell^2/2)~,
\end{equation}
where we used that $\omega_{\rm c} = 2\pi\hbar/(m^*A_{\rm M})$ in the final equality.

The dynamical dielectric function then equals
\begin{equation}
  \varepsilon({\bm q},\omega) = 1 - \frac{2\pi e^2}{q \bar{\varepsilon}} \chi_0({\bm q},\omega)
    = 1 + \frac{2\pi n_0 e^2 q}{m^* \bar{\varepsilon}} \frac{1}{\omega_{\rm c}^2 - (\omega + i\eta)^2} \exp(-q^2 \ell^2/2)~,
\end{equation}
where, in the last equality, we used that $n_0=1/A_{\rm M}$. Since plasmons are defined by the roots of the dielectric function, we find that their dispersion is given by
\begin{equation}  \label{eqn:magnetoplasmon-extra-app}
  \omega_{\rm pl}(q) = \sqrt{\omega_{\rm c}^2 + \omega_{\rm 2D}^2(q) \exp(-q^2 \ell^2/2) }~,
\end{equation}
with $\omega_{\rm 2D}(q) = \sqrt{2\pi n_0 e^2 q/(m^* \bar{\varepsilon})}$ as in the main text. In the limit ${\bm q} \to {\bm 0}$, Eq.~\eqref{eqn:magnetoplasmon-extra-app} reduces to the long-wavelength dispersion~\eqref{eqn:magnetoplasmon} that we previously obtained from the inter-band conductivity evaluated at the local level.

\end{document}